\documentclass[12pt]{elsarticle}

\usepackage{graphicx}
\usepackage{epsfig}
\usepackage{subcaption}
\usepackage{amssymb,amsmath}
\usepackage{float}
\usepackage{soul,xcolor}
\usepackage{todo}
\usepackage{hyperref}
\usepackage[percent]{overpic}

\renewcommand{\vec}[1]{\mathbf{#1}}

\begin{document}
\begin{frontmatter}

\title{Dynamic capillary phenomena using Incompressible SPH}

\author[iisc]{Prapanch Nair\corref{cor1}}
\ead{prapanch.nair@fau.de}

\author[iisc]{Thorsten P\"oschel\corref{cor2}}

\cortext[cor1]{Principal corresponding author}

\address[iisc]{ Institute for Multiscale Simulation, Friedrich-Alexander Universit\"at Erlangen-N\"urnberg, Erlangen.}

\begin{abstract}

Grid based fluid simulation methods are not able to monolithically  
capture complex non-linear dynamics like
the rupture of a dynamic liquid bridge
between freely colliding solids,
an exemplary scenario of capillary forces competing with 
inertial forces in engineering applications.  
We introduce a new Incompressible Smoothed Particle Hydrodynamics method for
simulating three dimensional fluid-solid interaction flows with capillary 
(wetting and surface tension) effects 
at free surfaces. This meshless approach presents 
significant advantages over grid based approaches in terms of being monolithic
and in handling interaction with free solids. 
The method is  validated for accuracy and stability in dynamic scenarios involving 
surface tension and wetting. We then present three dimensional simulations of 
crown forming instability following the splash of a liquid drop, and 
the rupture of a liquid bridge between two colliding solid spheres, to show 
the method's advantages in the study of dynamic
micromechanical phenomena involving capillary flows. 
\end{abstract}


\end{frontmatter}

\section{Introduction}
\label{intro}

Appreciation of non-linear micro mechanical phenomena is crucial to advance the 
efficiency of many production processes that are aided by the presence of a momentary 
liquid phase. Processes such as wet fluidized beds \cite{mikami1998numerical}, powder agglomeration \cite{hussain2013modeling},
pelletizing \cite{tsunazawa2016contact} and spray painting \cite{ye2005numerical} in industrial scales are essentially 
multiscale phenomena where the macroscopic behaviour is predicted by simulations
using simple analytical models for the smaller scale mechanics. Improving 
the understanding of micromechanics using well resolved simulations is crucial  
to improve the models employed in macroscopic simulations. Dynamic capillary 
scenarios in small scales are difficult to study experimentally and traditional 
fluid simulation methods have their limitations and therefore there is an increasing
interest in exploring meshless approaches. 

Traditional multiphase flow simulation methods employing the Volume of fluid (VOF) method \cite{hirt1981volume},
Constraint interpolation method (CIP) \cite{yabe2001constrained}, Level set or Coupled Level set VOF methods \cite{tomar2005numerical}
are increasingly used to simulate such phenomena. However, they have their limitations in 
handling
high density ratio or free surfaces \cite{popinet2009accurate}, moving three-phase contact line consistent 
with a no-slip wall \cite{renardy2001numerical,dussanv1976moving}  and interacting solids with 
all six degrees of freedom. While higher order consistency is easy to achieve in 
Eulerian methods, multi component simulations require coupling of different 
numerical approaches (for example, the Immersed Boundary Method \cite{peskin2002immersed}) limiting the ease of set up of these simulations.

Meshless Lagrangian
simulation methods, for example the Smoothed Particle Hydrodynamics (SPH) and its derivatives,
have the advantage of handling  complex shaped
free surfaces \cite{monaghan1994simulating}---a better approximation to a typical 
liquid-air system than a finite density ratio and explicit interactions with solids \cite{nair2014improved,nair2017study}. 
While free surfaces 
are implicitly handled by traditional SPH methods with 
a weak compressibility model, they require non-trivial treatment in case of 
the more 
accurate Incompressible SPH (ISPH) methods \cite{lee2008comparisons}, which solves for 
a pressure field for strict 
incompressibility. An accurate semi-analytic free surface model for ISPH was 
implemented recently \cite{nair2014improved}, expanding the scope of this 
more accurate version of SPH. 
Particularly, this approach allows finite pressure gradients close to the free surface 
making it possible to couple different interface tension models to the free
surface. 

Surface tension was initially implemented in SPH using a Continuum Surface 
Force (CSF) model\cite{brackbill1992continuum, morris2000simulating}, following 
Eulerian two phase flow simulation methods \cite{brackbill1992continuum}.
Since then CSF based surface tension models in SPH have improved considerably in 
accuracy and stability \cite{adami2010new,colagrossi2003numerical}. In CSF, the 
surface tension force is modeled as a volumetric force proportional to the 
interface curvature, which in turn is obtained by computation of divergence of 
a Heaviside step function \cite{morris2000simulating} 
or by geometrically
reconstructing the interface \cite{zhang2010simulation}. While the former is limited in its 
application to interfaces with fluid on both sides, the latter requires 
expensive computations to identify particles at the interface \cite{zhang2010simulation}. Inspired by the molecular origin of 
surface tension phenomena, a  model based on pair-wise forces was introduced 
by \cite{nugent2000liquid} for a van der Waals liquid drop. 
This approach was further improved in \cite{tartakovsky2005modeling} and applied to 
a number of free surface problems. Such approaches are also used in computer graphics 
research for emulating optically plausible behaviour with considerably less computational
cost \cite{akinci2013versatile}. In a recent work \cite{tartakovsky2016pairwise}, 
the relations between  macroscopic parameters like surface tension coefficient 
and apparent contact angle to the strength and nature of the  pairwise particle 
forces was elucidated. In this work, a number of validation
cases were presented by the authors establishing the method as an accurate 
surface tension model of engineering importance. 


In the present paper, we present a novel ISPH method that couples an accurate 
free surface model to a pairwise force capillary model to simulate dynamic capillary 
effects. The relation between inter particle force strengths and the macroscopic
parameters like surface tension and contact angle are given for free surfaces. 
Different aspects of the method such as dynamics due to surface tension,
contact angle, capillary force balance etc., are separately validated. 

We then apply the method to the simulation of two three dimensional problems 
exemplary of dynamic capillary effects. First, we observe the onset of instability 
and the breaking of symmetry following splash of a drop on a liquid
film. Second, we obtain the critical velocity for agglomeration of two 
colliding wet solids by observing the rupture of the liquid bridges as the 
solids depart and compare it with theoretical results.

\section{ISPH Formulation}
\label{sphformu}
The governing equation and SPH discretization used for incompressible fluids 
with free surface is presented here. The philosophy and basic formulation 
of the SPH method can be found in a number of works, for example \cite{violeau2012fluid}, 
and here we present only the SPH approximations that are 
relevant to the presented method.

\subsection{Governing Equations}
Momentum conservation equations for a Newtonian fluid are solved using the SPH 
method in an updated-Lagrangian frame of reference. By updated we mean that the 
reference configuration for application of constitutive relation of the material 
is updated with time, as opposed to a purely Lagrangian method where the reference
configuration is a relaxed initial state. The Navier--Stokes equations governing the 
momentum conservation of incompressible isothermal flow are given by,
\begin{equation}
\frac{d \vec{u}}{dt} = \frac{1}{\rho}\left(-\nabla p + \nabla \cdot \left(2 \mu \vec{D} \right) \right)+ \vec{f}^B ,
\label{ns-eqn}
\end{equation}
where $\vec{u}$ is the velocity, $p$ is the pressure, $\rho$ is the density, $\mu$ is the coefficient of viscosity of the fluid,   
      $\vec{D} = (\nabla\vec{u}+\nabla\vec{u}^T)/2$ is the deformation rate tensor, 
        $\vec{f}^B$ is the body force per unit mass on the fluid element 
        and $t$ is the time. The Navier-Stokes equation has been written in 
        the Lagrangian formulation and $d/dt$ denotes the material derivative.
The mass conservation equation for incompressible flows is given by,
\begin{equation}
\nabla \cdot \vec{u} = 0.
\label{continuity}
\end{equation}
\setstcolor{red}
The governing equations are discretized on a particle domain in SPH. As a model for
surface tension, a molecular dynamics inspired pairwise force model \cite{tartakovsky2016pairwise} is superimposed
on the particle system following the observation that molecular forces are superposable
on forces derived from momentum conservation equations on the same particle system \cite{nugent2000liquid}. 

\subsection{SPH formulation}

The SPH discretization of the governing equations \eqref{ns-eqn} together with a
pairwise force model \cite{tartakovsky2016pairwise} is as follows:
\begin{equation}
  \begin{split}                                                                 
   \left. \frac{d \vec{u} }{d t}\right| _a = & - \sum_b m_b \left(  \frac{p_a}{\rho_a^2}  +\frac{p_b}{\rho_b^2}\right) \nabla_a W_{ab}  \\
                                  &  + \sum_b m_b \left( \frac{2\mu }{(\rho_a \rho_b)}\frac{\vec{r}_{ab} \cdot \nabla W_{ab}}{r_{ab}^{2}+\epsilon^2}\right) \vec{u}_{ab} + \vec{f}^{\textrm{int}}_a +\vec{f}_a^b,
      \end{split}  
      \label{eq:sphdiscretization}
\end{equation}
where $m$ is the mass, $\rho$ is the density, $p$ is the pressure, at a particle 
identified by the subscript $a$ or its neighbor $b$. Here $\vec{r}_{ab}$ is the 
displacement vector between particles $a$ and $b$ and $r_{ab}$, its magnitude. 
The function $W$ is the symmetric and positive definite smoothing function, 
also known as the kernel for the SPH
discretization defined for a particle pair as $W_{ab} = W(r_{ab},h)$, where $h$ is 
the smoothing length of the kernel. The kernel has a compact support and its 
domain is cut off by a factor $q_\textrm{cutoff}$ times the smoothing length $h$
in space. The pairwise forces  $\vec{f}_{ab}^{\textrm{int}}$, applied to 
simulate interfacial force is explained in Sec. \ref{sec:pairwise} and $\vec{f}_a^b$
denotes the body force per unit mass acting on particle $a$ and represents the 
gravitational acceleration in the present work. We have used 
a viscous force approximation (the second term on the right hand side) that 
is extensively used in SPH literature \cite{morris1997modeling}:
\begin{equation}                                                                
   \nabla \cdot \left(\frac{\mu}{\rho} \nabla \vec{u}\right)_a = \sum_b         
   m_b\left( \frac{\mu_a + \mu_b}{\rho_a \rho_b}\frac{\vec{r}_{ab} \cdot        
   \nabla_a W_{ab}}{r_{ab}^{2}+\epsilon^2}\right)\vec{u}_{ab},                       
 \end{equation}
 where $\vec{u}_{ab}$ is the relative velocity vector between particles $a$ and $b$,
 $\epsilon$ is a small positive number (usually chosen as $(0.01h)^2$) to avoid division by zero
 for a rare incident of particles overlapping each other, and a pressure gradient approximation (the first term on the right hand side) applicable to multiphase flow problems  \cite{colagrossi2013smoothed, szewc2012analysis}:
  \begin{equation}                                                                
   \left(\frac{\nabla p}{\rho }\right)_{a} = \sum_b m_b                          
   \frac{p_a + p_b}{\rho_a \rho_b} \nabla_a W_{ab}.                               
   \label{multiphase_gradient}                                                   
\end{equation} 
Near solid walls, the gradient and divergence approximations require a filled
kernel neighborhood. This is achieved by distributing static particles 
along solid walls with the same particle spacing as in the initial spacing of 
fluid particles. These wall particles can be modelled as
belonging to rigid bodies to simulate free solids interacting with the liquid. Such
an approach conserves linear and angular momentum and can be seen by a  
force balance across the solid-liquid interface as explained in \cite{nair2014improved}.

At free surfaces, standard weakly compressible  SPH is known to naturally 
satisfy a zero pressure Dirichlet boundary condition corresponding to a moving  
interface \cite{monaghan1994simulating} if a conservative pressure gradient 
(for example, eq. \ref{multiphase_gradient}) approximation is used. 
However, explicit application of Dirichlet boundary condition is necessary 
if a pressure solver 
is invoked to compute the pressure field. It is important for the Dirichlet 
boundary condition to be applied in a consistent and accurate manner to 
preserve the accuracy of ISPH.

\subsection{ISPH and the free surface formulation}

Following grid based methods for incompressible flows, where a divergence free
constraint is imposed on velocity field, ISPH solves for the pressure Poisson equation 
\begin{equation}
  \nabla \cdot \left( \frac{\nabla p}{\rho} \right) = - \frac{\nabla \cdot \vec{u}}{\Delta t},
    \label{eq:ppe}
\end{equation}
on the particle domain. The SPH discretization of this equation \cite{lee2008comparisons} is:
\begin{equation}
    \left.  \nabla \cdot \left( \frac{\nabla p}{\rho}\right)\right|_a =\sum_{b } \frac{m_b}{\rho_b}\frac{4}{\rho_a +\rho_b}\left(P_a - P_b\right)\frac{\vec{r}_{ab} \cdot        
   \nabla_a W_{ab}}{r_{ab}^{2}+\epsilon^2}.
      \label{laplacian_p}                                                         
    \end{equation}   
When a linear system is solved on the particle
domain to obtain pressure, application of Dirichlet boundary condition for pressure 
is necessary. A semi-analytic model was introduced in a previous work \cite{nair2014improved}  to 
implement free surface Dirichlet boundary condition with greater accuracy and robustness over
previous approaches involving identification of free surface particles \cite{lee2008comparisons,
violeau2016smoothed}.
This model is implemented by modifying the linear system for solving the pressure Poisson equation 
in the SPH particle domain as:
 \begin{equation}                                                 
   \begin{split}
     \left. \nabla \cdot \left( \frac{\nabla p}{\rho}\right)\right|_a = & (p_a - p_o)  \kappa - \sum_{b_i} \frac{m_b}{\rho_b}\frac{4}{\rho_a +\rho_b}p_b \frac{\vec{r}_{ab} \cdot        
   \nabla_a W_{ab}}{r_{ab}^{2}+\epsilon^2} \\
   & + \sum_{b_i}\frac{m_b}{\rho_b}\frac{4}{\rho_a +\rho_b}p_o\frac{\vec{r}_{ab} \cdot        
   \nabla_a W_{ab}}{r_{ab}^{2}+\epsilon^2}               
 \end{split}
               \label{freesurf_eq}                                                            
              \end{equation}
   where $p_0$ represents the ambient pressure and  $\kappa $, given by    
   \begin{equation}                                                             
     \kappa = \sum_b \frac{m_b}{\rho_b} \frac{4}{\rho_a +\rho_b}\frac{\vec{r}_{ab} \cdot        
   \nabla_a W_{ab}}{r_{ab}^{2}+\epsilon^2}  ,
   \end{equation}
   represents a kernel completion factor. Here,  The derivation of both the above equations 
   can be seen in \cite{nair2014improved}. In effect, this model simply modifies the 
   terms in the leading diagonal of the linear system for pressure.
The time integration of the field variables are performed by a velocity Verlet 
integration algorithm \cite{allen1989computer} given by: 
\begin{align}
  \vec{r}_a(t+\Delta t) &= \vec{r}_a(t) +\Delta t \vec{u}_a (t) + \frac{\Delta t^2}{2}\vec{f}^\text{tot}_a (t) \\
  \vec{u}_a(t+\Delta t) &= \vec{u}_a(t) + \frac{\Delta t}{2}\left( \vec{f}^\text{tot}_a (t)+\vec{f}^\text{tot}_a (t+ \Delta t)\right) .
\end{align}
Here, $\vec{f}^\text{tot}_a$ is the total acceleration of a particle $a$ contributed by
both continuum (viscous, pressure, body forces) and inter-particle forces. The integration 
time step $\Delta t$ is set to satisfy 
the following constraint for numerical stability \cite{tartakovsky2016pairwise}:
\begin{equation}
  \Delta t \leq \min_a \left( 0.25\frac{h}{3|\vec{u}_a| } , 0.25\sqrt{\frac{m_a h}{3|\vec{f}^\text{int}_a|}}, 0.25\frac{\rho h^2}{9\mu} \right).
\end{equation}

\done\todo{time integration and algorithm }
\subsection{Pairwise-force model for free surface ISPH}
\label{sec:pairwise}
The pairwise force required to model capillary effects 
needs to be  repulsive in 
the short range and attractive in the long range and may be much less stiffer than
the potentials used in molecular dynamics simulations \cite{rowlinson2013molecular}.
The inter-particle acceleration term that appears in eq. \ref{eq:sphdiscretization} is 
given by:
\begin{equation}
  \vec{f}_a^{\textrm{int}} = \sum\limits_{b} \frac{ F_{\alpha\beta}^{\textrm{int}}({r}_{ab})}{m_a} \frac{\vec{r}_{ab}}{r_{ab}}, 
\end{equation}
where $\alpha$ and $\beta$ represent the phases of the particles $a$ and $b$, respectively, separated
by a vector $\vec{r}_{ab}$ in space.  
The pairwise force magnitude as a function of displacement between particles 
is given as \cite{tartakovsky2005modeling}:

\begin{equation}
  F_{\alpha\beta}^{\textrm{int}} (r_{ab}) =\begin{cases}
    -s_{\alpha\beta}\cos \left( \frac{3\pi}{4} q_{ab} \right) & q_{ab} = \frac{r_{ab}}{h} \leq q_\textrm{cutoff} \\
    0 & q_{ab} = \frac{r_{ab}}{h} > q_\textrm{cutoff},
  \end{cases}
  \label{eq:interaction}
\end{equation}
where $q_\textrm{cutoff}$ is the ratio of cut off length of the kernel to the smoothing length $h$. Here,
we have chosen the cut off length of the kernel for SPH computation to be the same for
the inter-particle forces. Also, $s_{\alpha\beta}$ is the interaction strength between particles of 
phases $\alpha$ and $\beta$ 
respectively. Symmetry in the strength ensures conservation of linear momentum between
particles, ensuring conservation of linear momentum in the SPH discretization.
The pairwise force based SPH model is becoming increasingly popular 
\cite{tartakovsky2005modeling,liu2010smoothed,tartakovsky2006pore} in literature owing
to its ease of application and robustness. In a recent work \cite{tartakovsky2016pairwise} 
a detailed explanation on how the macroscopic parameters such as surface tension,
 contact angle and the pressure resulting from pairwise forces, called the `virial pressure' 
can be related to the strength of the pairwise forces is provided. This approach 
was also applied to free surface
flows in earlier works \cite{nugent2000liquid, tartakovsky2005modeling}. 
However the application has been limited to weakly compressible
SPH solvers, since ISPH free surface flows . In the present work the pairwise force model is 
applied to the ISPH algorithm with the free surface treatment presented in the previous
section. 

The stress at a given point in a particle system with given pairwise forces can be 
computed according to the Hardy's formula \cite{rowlinson2013molecular}.
This formula is given as a sum of stress due to inter-particle forces and the convection of 
particles. In the present work, since a hydrodynamic governing equation is used
to model the fluid, only the stress due to inter-particle forces becomes relevant,
and is given by:
\begin{equation}
  \vec{T}_{\textrm{int}}(x) = \frac{1}{2}\sum\limits_{a=1}^{N}\sum\limits_{b=1}^{N} \vec{f}_{ab} \otimes (\vec{r}_a -\vec{r}_b)\int\limits_0^1\tilde{\psi }_\eta (\vec{x} -s\vec{r}_a -(1-s)\vec{r}_b)ds,
  \label{hardys}
\end{equation}
where the function $\tilde{\psi}$ is a weighting function with a compact support $\eta$ which should be greater the SPH smoothing length
$h$. Following \cite{tartakovsky2016pairwise},  $\psi_\eta$ is chosen to be the 
product of constant valued functions with compact support $\eta$ , in each spatial dimension. 
Thus, 
\begin{equation}
  \tilde{\psi}_\eta(\vec{x}) = \prod_l^\nu \psi_\eta (x_l) ,
\end{equation}
where $\nu$ is the spatial dimension, and 
\begin{equation}
\psi_\eta (x_l) = \frac{1}{\eta}\psi\left( \frac{x}{\eta} \right), \quad \psi(x_l) =\begin{cases}
  1 & \textrm{if } x_l \in (-1/2,1/2), \\
  0 & \textrm{otherwise}.
\end{cases}
\end{equation}
It is convenient to set the cut off radius for both the pairwise forces and the 
SPH kernel to the same value. In this work we have have uniformly set the cut off to 
be two times the smoothing length, $q_\textrm{cutoff} = 2$.  
The surface tension at the interface between two phases $\alpha $ and $\beta$ can be
obtained by integration of the tangential stress components along a coordinate, say $z$, perpendicular
to the interface as
\begin{equation}
  \sigma_{\alpha\beta} = \int\limits_{-\infty}^{+\infty} [ T_\tau (z) - T_n(z)] dz.
\end{equation}
Here, $T_\tau (z)$ and $T_n (z)$ are the tangential and normal components of the stress 
when the $z$ coordinate is  perpendicular to the interface. Invoking eq. \ref{hardys} we write
\begin{equation}
  \sigma_{\alpha\beta} = \mathcal{T}_{\alpha \alpha} + \mathcal{T}_{\beta \beta} -2\mathcal{T}_{\alpha \beta},
\end{equation}

where the integrals of tangential stress components due to interaction force between 
particles of phases $\alpha - \alpha$, $\beta - \beta $ and $\alpha-\beta$ are represented by 
$\mathcal{T}_{\alpha\alpha}$, $\mathcal{T}_{\beta \beta} $ and $\mathcal{T}_{\alpha \beta}$ respectively. 
For a given smoothing kernel $\phi(r)$, these components are derived in the appendix
of \cite{tartakovsky2016pairwise}. For the free surface problems that are of interest 
to the present work, where only one liquid phase ($\alpha = l $) is present, the surface tension 
can be related to the pairwise force as follows for 2D and 3D cases:
\begin{align}
  \mathcal{T}_{ll}^{(\textrm{2D})} &= -\frac{1}{6}\pi \left(\frac{\rho_l}{m_l}\right)^2\int\limits_0^\infty z^3F^{\textrm{int}}_{ll}(z) dz,  \\
  \label{ststrength}
  \mathcal{T}_{ll}^{(\textrm{3D})} &= -\frac{1}{16}\pi \left(\frac{\rho_l}{m_l}\right)^2\int\limits_0^\infty z^4F^{\textrm{int}}_{ll}(z) dz .  
\end{align}
where  $z$ is the coordinate in the direction perpendicular to the interface, and $m_l$ and $\rho_l$ are the mass and density of particles representing the phase, $l$. 

One important assumption being made in 
 the above derivations is that the stresses are integrated across a plane interface, which
 in effect amounts to having a radius of curvature much larger than the smoothing length. 
 For the specific piece-wise force function we have presented, the relation between strength of the 
 force between two liquid particles $s_{11}$ and the surface tension coefficient $\sigma$ can be derived as:
 \begin{align}
   \sigma &= \lambda s_{ll} h_\text{r}^4 , \quad \textrm{for 2D and}  \\ 
   \sigma &= \lambda s_{ll} \frac{h_\text{r}^5}{\Delta x}  \quad \textrm{for 3D,}
   \label{eq:3dsurftens}
 \end{align}
 respectively.  Here, $h_\text{r}$ is the ratio of the actual smoothing length 
of the kernel to the initial particle spacing $\Delta x$ (here we use a square lattice arrangement of
 particles). Note that these expressions correspond to the specific 
 choice of pairwise force function and compact support. The constant due to 
 integration of the pairwise force function, $\lambda$ takes the value 
 $0.0476$ in 2 dimensions and $0.0135\pi$ in 3 dimensions, respectively,
 for the interaction function given by eq. \ref{eq:interaction}.  Note  the occurrence of 
 the absolute value of initial particle spacing in the three 
 dimensional version of this relation.

The pairwise force model also introduces an artificial virial pressure. This pressure
can be computed for a given particle configuration and pairwise force function from the following equation:

\begin{align}
  p^v_l &= -\frac{1}{2} \pi \left(\frac{\rho_l}{m_l}\right) \int \limits_0^\infty z^2 F^{\text{int}}_{ll} (z) dz \\
  p^v_l &= -\frac{2}{3} \pi \left(\frac{\rho_l}{m_l}\right) \int \limits_0^\infty z^3 F^{\text{int}}_{ll} (z) dz 
  \label{virialeq}
\end{align}
 for two and three dimensions, respectively. For the incompressible flows that are
 of interest to the present work, this pressure is additive to the hydrodynamic 
 pressure computed to satisfy incompressibility. Hence, this pressure can be computed
 and subtracted from the pressure obtained from solving the linear system given by eq. \ref{laplacian_p}.
The contact angle made by the liquid with a solid substrate can then be controlled 
by appropriate ratio of pairwise force strength between particles of different phases. 
In the present scenario liquid and solid phase alone are considered. The contact
angle can be computed from a surface energy balance for surface energy between free surface 
and solid liquid interface. The contact angle is given by \cite{tartakovsky2016pairwise}:

\begin{equation}
  \cos \theta_0 = \frac{-s_{ll} + 2s_{l s}}{s_{ll}},
    \label{eq:ca}
\end{equation}
where, $\theta_0 $ is the contact angle made by the liquid free surface with the 
solid substrate, $s_{ll}$ and $s_{ls}$ ($\alpha =l$ and $\beta = s$) are the strengths of the pairwise force for 
liquid-liquid particle pairs and liquid-solid particle pairs respectively. The above 
equation is a simplification of the contact angle expression given in eq. 60 of  
\cite{tartakovsky2016pairwise}.

Smoothed Particle Hydrodynamics approximation of the momentum equation is similar
in form to a particle system with inter-particle forces explicitly defined between 
particles. Using the concept of composite particles,  \cite{violeau2012fluid} shows that 
a pressure force term can be derived similar in form to SPH pressure gradient, which 
also implies that a different set of inter-particle forces can be superimposed 
on the SPH particle system consistently. For ISPH this may be even more relevant
since the pressure field is smoother in general than weakly compressible SPH methods \cite{lee2008comparisons}.

\section{Validation and Results}
\label{results}
The above introduced capillary model based on pairwise forces applied at free surfaces coupled 
by the dynamics simulated by ISPH is a novel method and therefore requires 
careful validation before application 
to realistic scenarios. We first present validations of the above described ISPH free surface capillary model using 2D and 3D simulations. 
We solve dynamic
test cases where the absolute value of pressure modified by the 
presence of pairwise forces (virial pressure as in eq. \ref{virialeq}) is unimportant,
as the dynamics would be determined by the pressure gradient on a constant
density domain. Oscillating drop
test cases are presented in 2 and 3 dimensions to validate the macroscopic surface 
tension coefficient against the strength of the pairwise potential. Contact angles 
are measured at steady state and during transient
states of relaxation of a sessile droplet on a plane surface. 
Capillary rise of liquid through a capillary 
tube is also simulated to check for the model's capability to handle 
surface tension and contact angles in the same domain. We then use the 
algorithm to simulate the impact of a drop of liquid on a liquid film
in order to observe the onset of instability and the  breakage of axisymmety leading to the
 formation of a splash crown in three dimensions. Finally, together with a solid 
 interaction algorithm, we use the method to simulate rupture/sustenance of liquid bridges 
 following collision of solid spheres of industrially relevant dimensions with wet spots, 
 in order to demonstrate the promise of the algorithm in handling arbitrary geometries.
We use the Wendland kernel Ref.\cite{wendland1995piecewise} for all the test cases presented here, 
owing to its superior numerical stability properties \cite{dehnen2012improving, szewc2012analysis}.
In sections \ref{sec:st} and \ref{sec:ca}, we have not provided specific units
in our plots, since arbitrary units could be used without changing the results.
\subsection{Oscillation of a liquid drop}
\label{sec:st}

\begin{figure}[htb]
  \begin{subfigure}[b]{0.48\textwidth}
  \includegraphics[width=\textwidth]{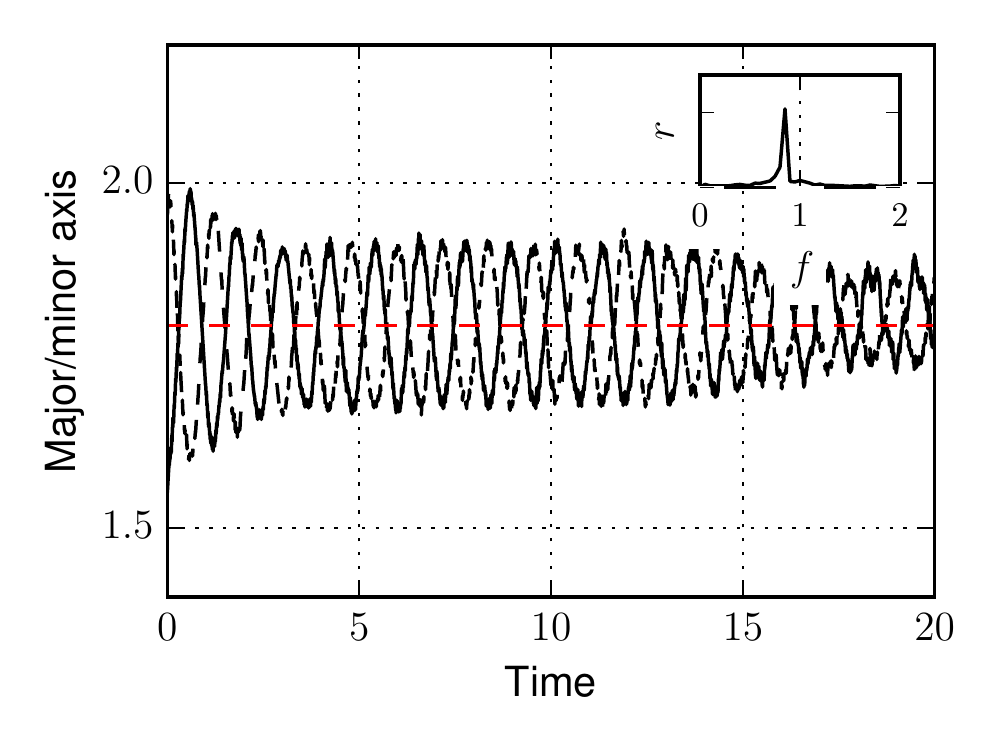}
  \caption{2D}
  \label{fig:osc2d}
\end{subfigure}
\begin{subfigure}[b]{0.48\textwidth}
  \includegraphics[width=\textwidth]{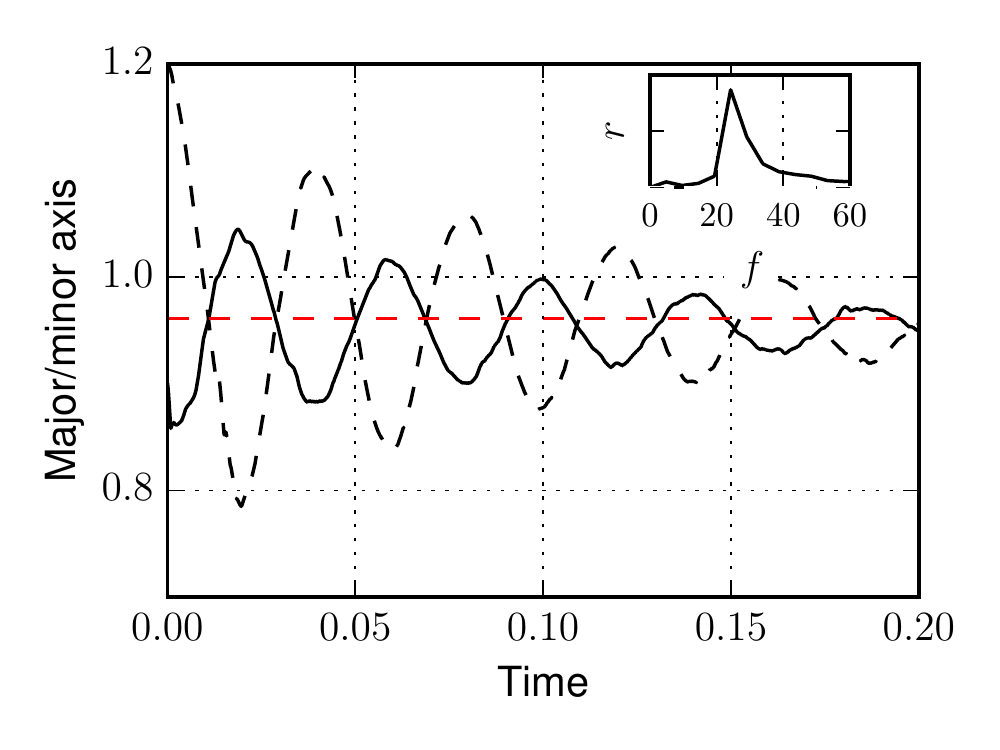}
  \caption{3D}
  \label{fig:osc2d}
\end{subfigure}
  \caption{Time response of radii of an oscillating drop about its 
  reference configuration in 2D and 3D. Insets show the frequency domain of
oscillation, showing the frequency of oscillation.}
\label{fig:osc}
\end{figure}

\begin{figure}[htb]
  \begin{subfigure}[b]{0.48\textwidth}
  \includegraphics[width=\textwidth]{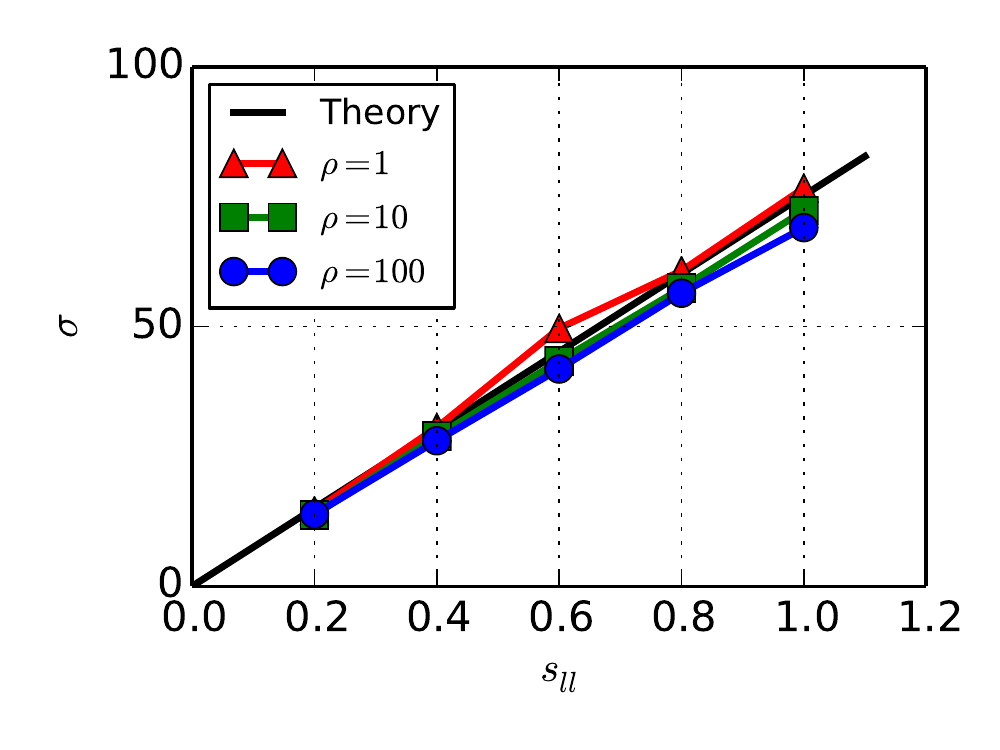}
  \caption{2D}
  \label{densityvariation}
\end{subfigure}
\begin{subfigure}[b]{0.48\textwidth}
  \includegraphics[width=\textwidth]{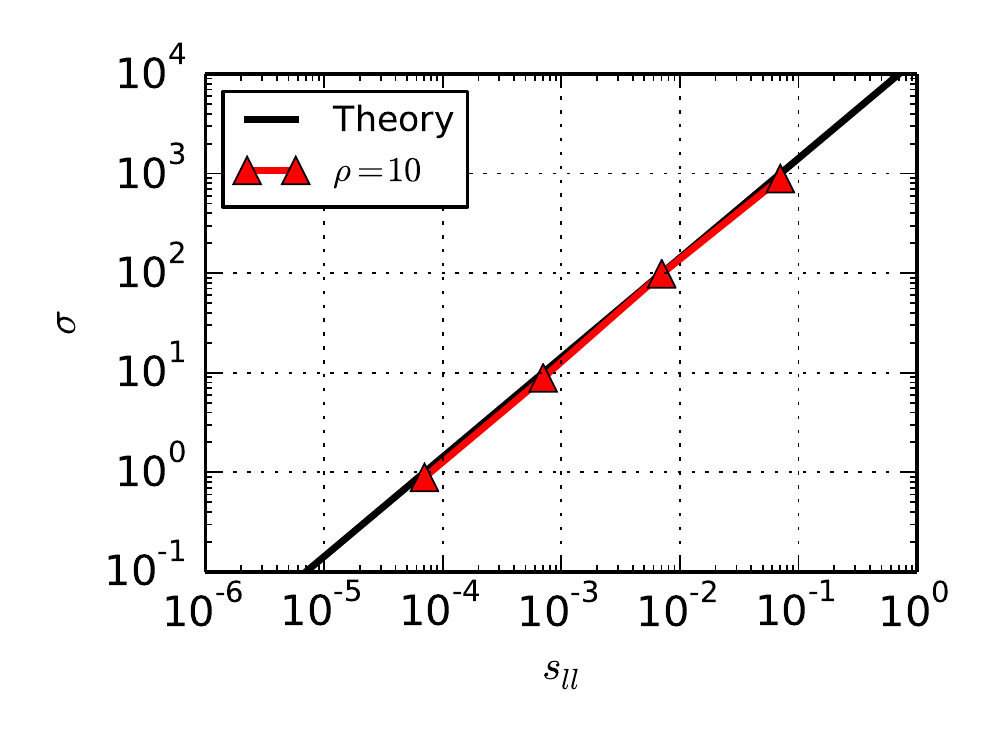}
  \caption{3D}
  \label{logvariation}
\end{subfigure}
\caption{Linear variation of surface tension coefficient with strength of 
pairwise potential. The solid black line is obtained from equation \ref{ststrength}.  
For the 2D cases, different densities were considered for the oscillating drop, and
potential strengths across orders of magnitude  were used in the 3D cases.
}
  \label{stcoefficient}
\end{figure}
\begin{figure}[htb!]
  \includegraphics[width=\textwidth]{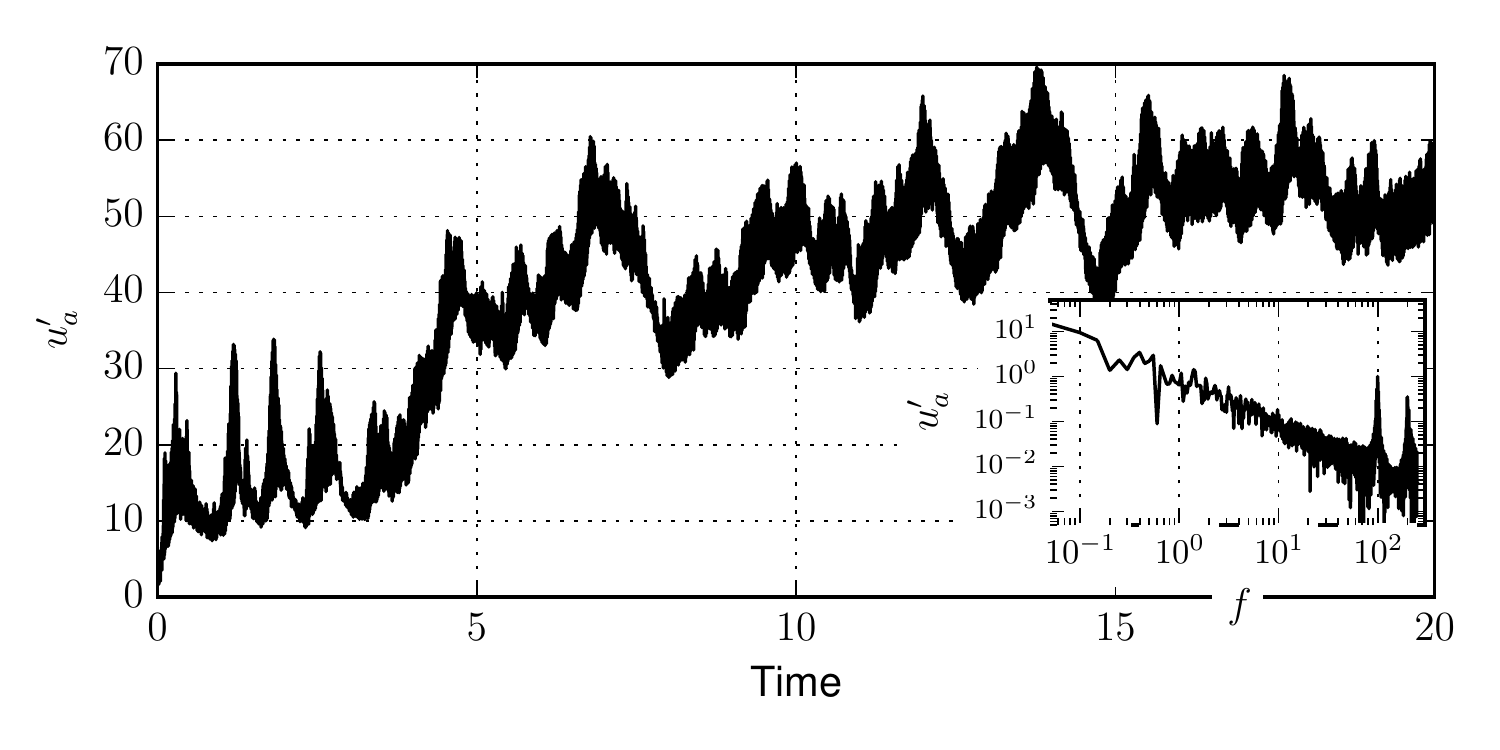}
  \caption{Variation of velocity fluctuation (eq. \ref{sph_temp}) of a particle with time. 
    Inset shows the presence of high frequency components in the 
    frequency domain of the signal, due to the superimposed 
  inter-particle forces.}
  \label{temperature}
\end{figure}

The surface tension coefficient for a given particle configuration 
can be verified by computing the frequency of oscillation of a drop of 
fluid about its circular (in 2D) or spherical (in 3D) minimal area 
reference geometry. The time period of oscillation for an inviscid infinite cylindrical jet is given by
\cite{nugent2000liquid} based on the general results of Rayleigh \cite{rayleigh1879capillary} as:
\begin{equation}
  T = 2\pi \sqrt{\frac{R^3 \rho}{6\sigma}}.
  \label{rayleigh2d}
\end{equation}
For three dimensions, the frequency of shape oscillation (second mode) of a drop is given by:
\begin{equation}
  T = \pi \sqrt{\frac{R^3 \rho}{2\sigma}},
  \label{rayleigh3d}
\end{equation}
where $T$ is the time period of oscillation, $\rho$ is the density of the liquid,
$R$ is the radius of the drop and 
$\sigma$ is the surface tension coefficient.

Surface tension can thus be measured from these equations by measuring the time 
period of oscillation of a drop, initially perturbed to an ellipse 
(or ellipsoid) of equal volume. 
The time response of the radius of the infinite 2D cylinder and sphere 
are shown in Fig. \ref{fig:osc}.
The simulation experiment is repeated with liquid drops of different densities 
and pairwise force strengths for 2D and these results are plotted in Fig. \ref{densityvariation}, and
in 3D (Fig. \ref{logvariation}) this linear relation is shown to hold good across orders of magnitude of pairwise force strengths. 
The results show that the relation between potential strength and surface tension is 
indeed linear, as seen in eq. \ref{eq:3dsurftens}. In these 
validation cases the smoothing length $h$ is taken to be $3.75$ times the initial 
particle spacing. A circle of radius $1$ unit is used with a resolution such that 
the radius is more than $5h$ in these simulations. No viscous model is used in this 
case. 

Though the oscillations are simulated well, the oscillations damp considerably 
with time, and is more pronounced in the 3D case. This damping is due to the 
velocity fluctuations in the particles due to the 
inter particle force model and warrants separate attention. In Fig. \ref{temperature} 
we have shown the particle velocity fluctuation for the 2D case, computed by subtracting velocity
of each particle from the SPH averaged velocity at that particle position as:
\begin{equation}
  u'_a = \sqrt{\left( u_{ia} - \sum_b u_{ib}W_{ab}\frac{m_b}{\rho_b} \right)^2},
  \label{sph_temp}
\end{equation}
where $u_{ia}$ is the velocity component of a particle $a$ in direction $i$, $b$ 
represent the neighbors of $a$ and 
the index $i$ is given in Einstein's summation notation. 
The inset in the plot shows the frequency domain of the fluctuation. Evidently,
there are remarkable peaks in an otherwise power law frequency distribution, at high frequencies.
Also, in fig. \ref{fig:osc} there is an initial drop in amplitude following 
which the amplitude remains constant for many oscillations before it deteriorates 
again. The initial damping could be due to the initial arrangement
of particles in an array, which quickly assumes a much lower energy configuration. 
Further damping is of course due to additional viscosity due to 
the high frequency oscillations of particles. All this warrants a separate 
study on the free energy of SPH particles when superimposed by an inter-particle 
force model. However as we seen here and in the following test cases, the ability of 
the method to simulate surface tension effects for many practical 
applications is quite promising.

\subsection{Validation of the contact angle model}
\label{sec:ca}
\begin{figure}[htb]
  \begin{subfigure}[b]{0.48\textwidth}
    \includegraphics[width=\textwidth]{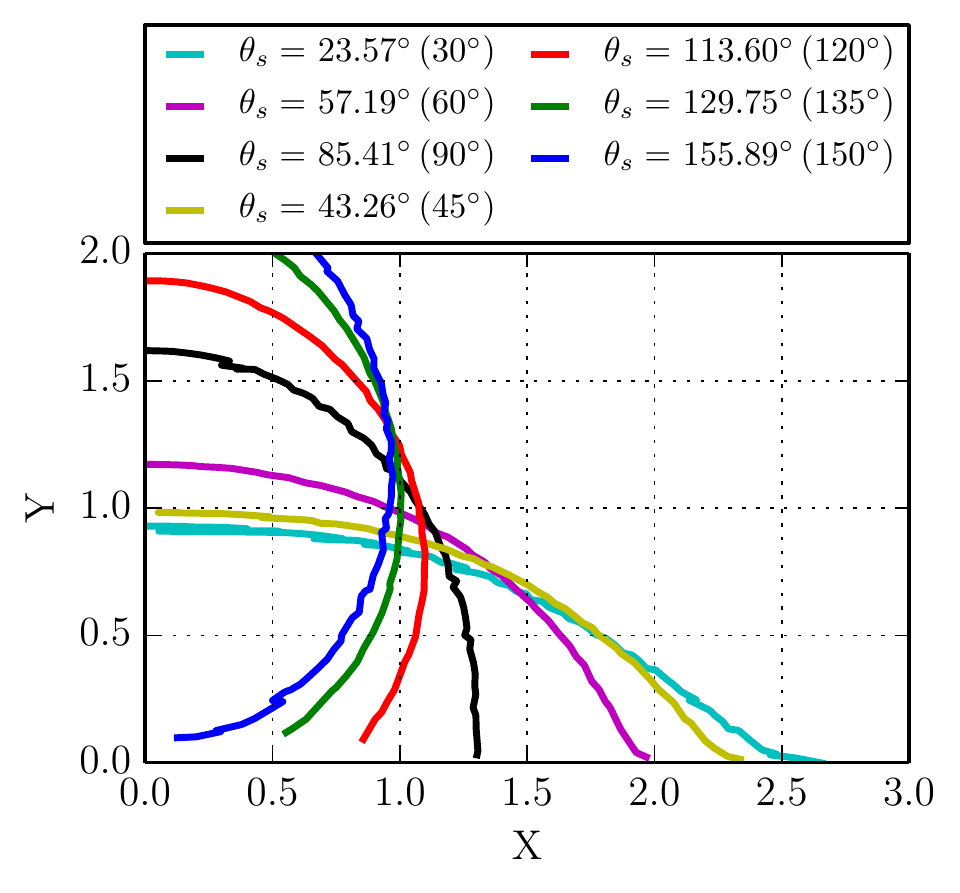}
  \caption{2D }
  \label{fig:ca2d}
  \end{subfigure}
  \begin{subfigure}[b]{0.48\textwidth}
    \includegraphics[width=\textwidth]{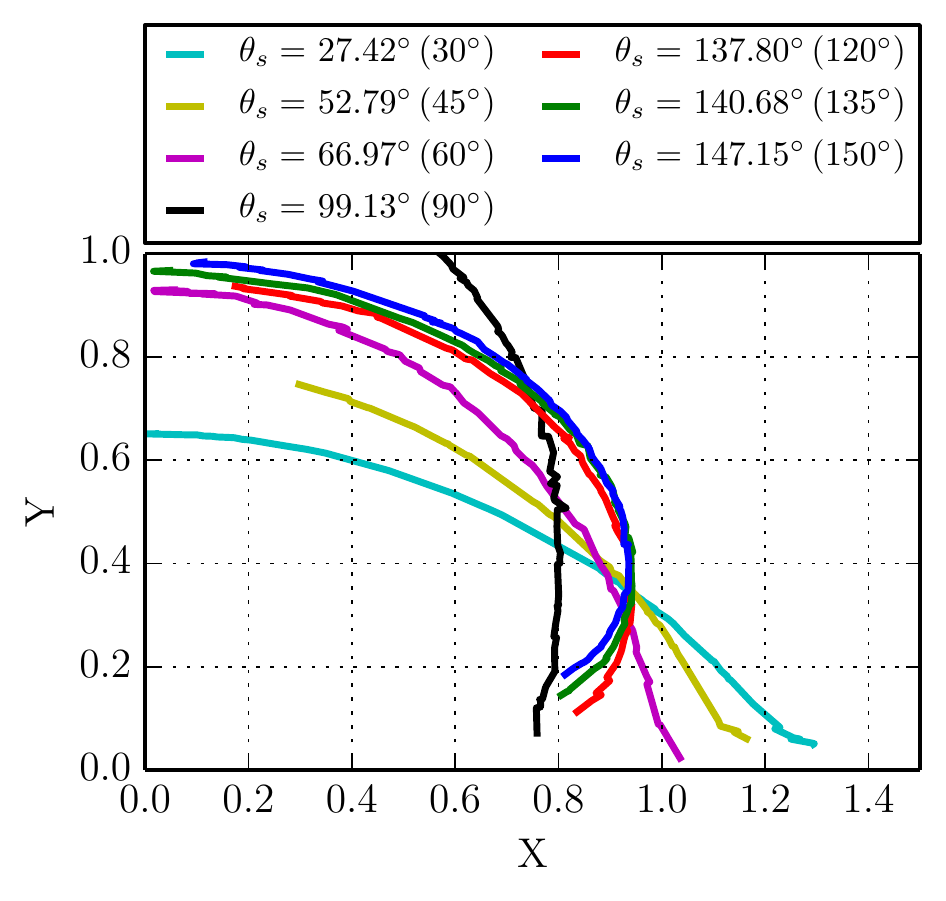}
  \caption{3D }
  \label{fig:ca3d}
  \end{subfigure}
  \caption{Contact angles of a sessile drop measured for different potential strength ratios. 
  Values in brackets are the expected angles according to eq. \ref{eq:ca} }
\end{figure}
\begin{figure}[htb]
  \includegraphics[width=5in]{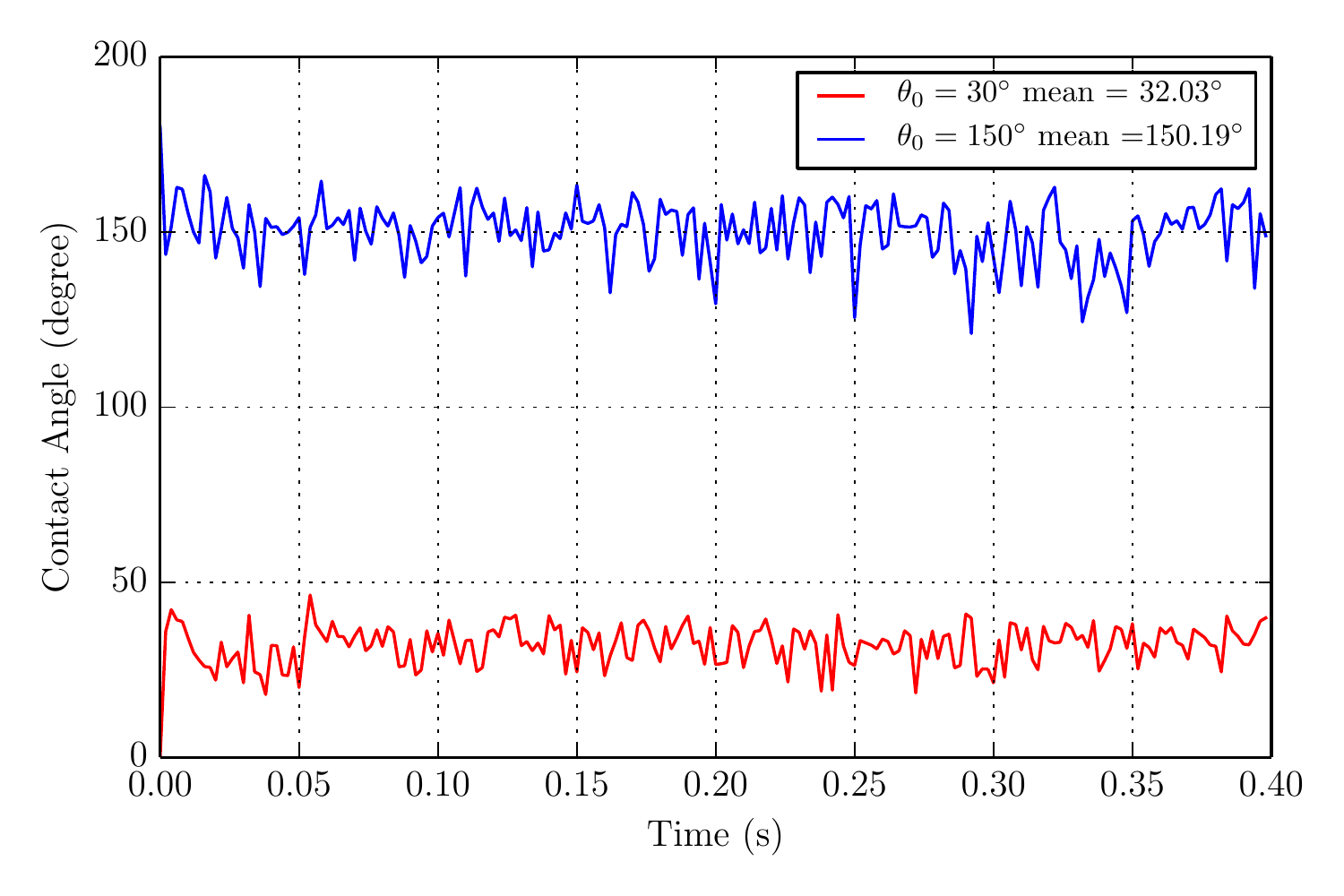}
  \caption{Time variation of the contact angle of a 3D sessile drop for the case of acute ($\theta= 30^\circ$) and obtuse ($\theta=60^\circ$) contact angles. The droplets were initiated from the geometry of hemisphere resting on a solid surface. }
  \label{fig:catime}
\end{figure}

The contact angle and incidental wetting behaviour can be controlled in the pairwise force 
ISPH by varying the relative potential strengths between particles of different phases
according to eq. \ref{eq:ca}. A droplet (in 2D and 3D) initially in a hemispherical configuration (semicircle in 2D) is placed on a solid
substrate and is allowed to relax. In the case of obtuse contact angles a gravitational
body force was applied on the particle to prevent it from bouncing off. After 
steady state was reached, a layer of particles lying on the free surface were obtained by 
a kernel summation criteria and first six particles close to the contact point 
were chosen and regressed linearly to obtain the apparent contact angle $\theta_s$. 
Figure \ref{fig:ca2d} 
and \ref{fig:ca3d} show the profile of the drop for different potential strength 
pairs. The bracketed values show the expected contact angles computed by eq. \ref{eq:ca} in this figure. The resolution and geometry of the drop is similar to the previous validation
case for oscillating drops.

In order to demonstrate the evolution of the contact angle with time, 
we show in fig. \ref{fig:catime}, 
the time variation of the contact angle (measured similar to the steady state contact angles) for an obtuse and an acute angle 
case from the 3D simulations. The plot shows that the angle remains 
reasonably steady with minor oscillations about its expected value during the relaxation
of the drop. This is comparable with pinning a contact angle heuristically applied in 
case of a traditional mesh based CFD method with a sharp interface model such 
as the volume of fluids (VOF) method. 

\subsection{Capillary Rise in 2D}
\begin{table}
  \begin{center}
\begin{tabular}{l|l|l}
  & \multicolumn{2}{c}{ Height }\\
  \cline{2-3}
  Width (mm) & Analytical & SPH \\
  \hline
  0.50 & 5.01 & 5.12  \\
  0.75 & 2.78 & 2.95 \\
  1.00 & 1.56 & 1.31\\ 
\end{tabular}
\caption{The capillary rise height for different capillary tube diameters (2D)}
\label{tab:capillary}
  \end{center}
\end{table}
\begin{figure}[htb]
  \begin{subfigure}[b]{0.32\textwidth}
    \includegraphics[width=\textwidth]{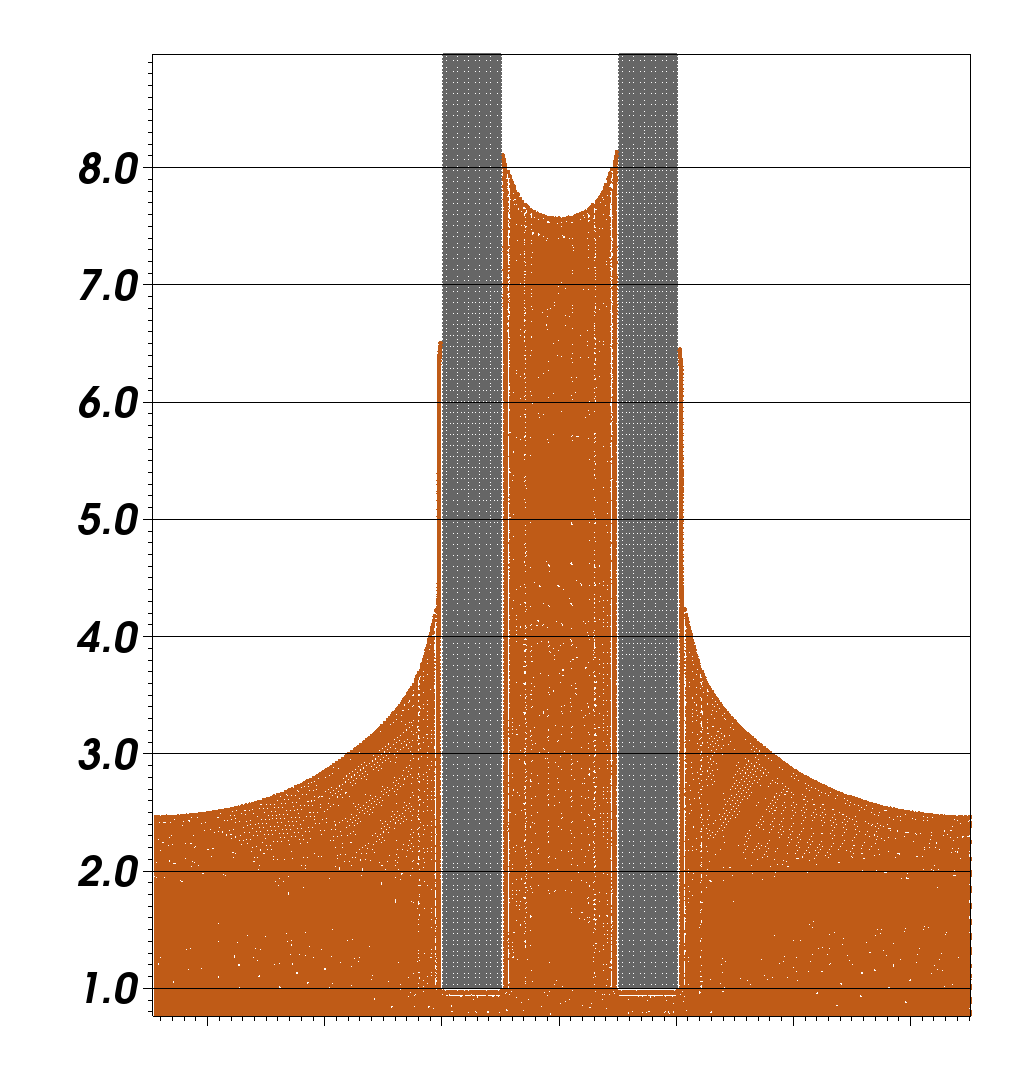}
  \caption{width$=0.5\,mm$}
  \end{subfigure}
  \begin{subfigure}[b]{0.32\textwidth}
    \includegraphics[width=\textwidth]{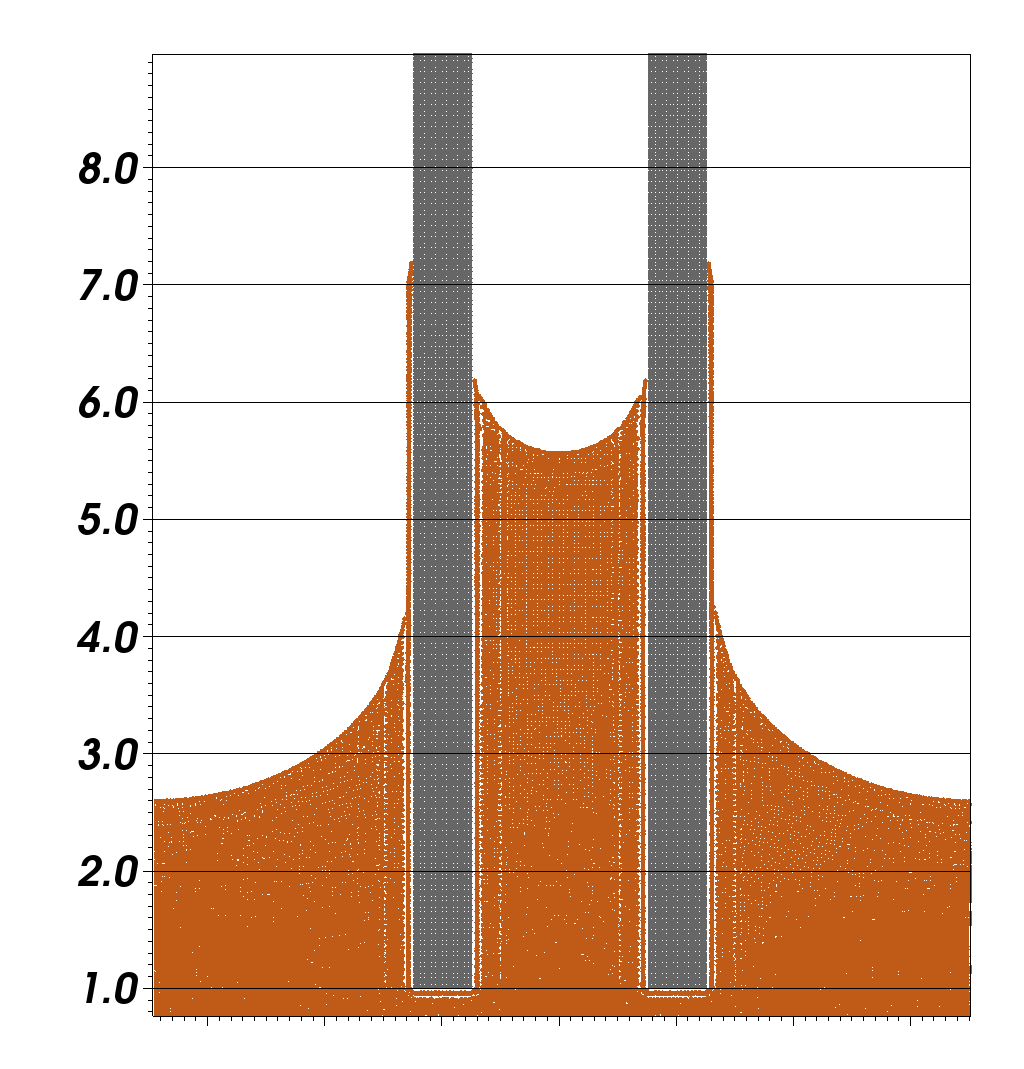}
  \caption{width$=0.75\,mm$}
  \end{subfigure}
  \begin{subfigure}[b]{0.32\textwidth}
    \includegraphics[width=\textwidth]{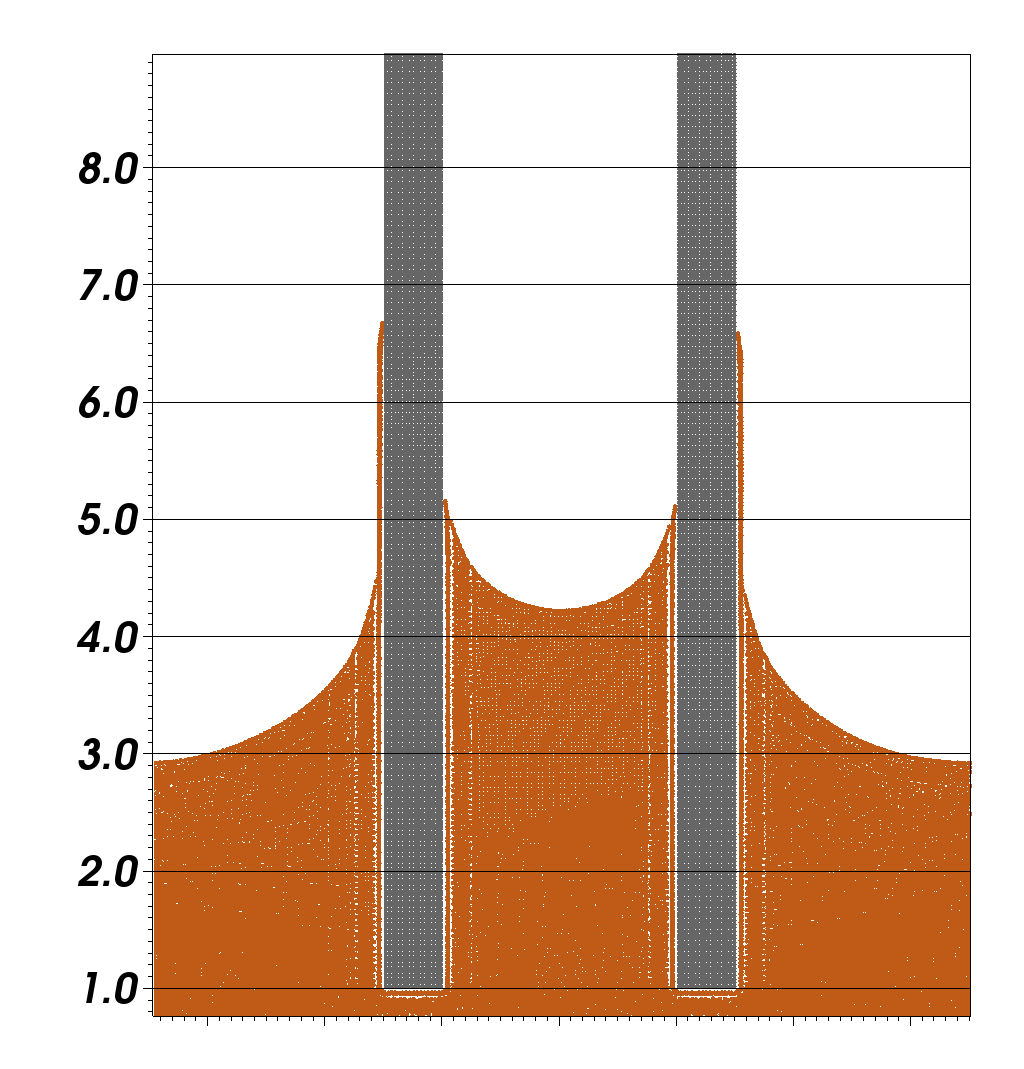}
  \caption{width$=1.0\,mm$}
  \end{subfigure}
  \caption{Capillary rise for different tube diameters}
  \label{fig:carise}
\end{figure}
Capillary rise of a liquid through a capillary tube is an intricate and useful
phenomenon for which analytical solution is straightforward. We construct
a simple 2D domain periodic in the horizontal direction with a capillary tube
inserted into it. A contact angle of $30^\circ $ is chosen in these test case, and 
is performed for different tube diameters. 

The height of capillary liquid column for different capillary diameters can be found in table \ref{tab:capillary}
and the steady state of the simulation for these cases can be seen in fig. \ref{fig:carise}. The 
method  predicts the rise of the liquid column reasonably accurately, giving 
confidence for simulations involving surface tension and wetting and their combined
effects.

\subsection{Drop impact on liquid film }
\begin{figure}[htb]
  \begin{subfigure}[b]{0.32\textwidth}
    \begin{overpic}[width=\textwidth]{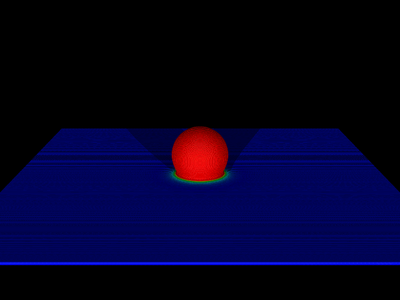}
           \put(3,52){\includegraphics[width=0.9\textwidth]{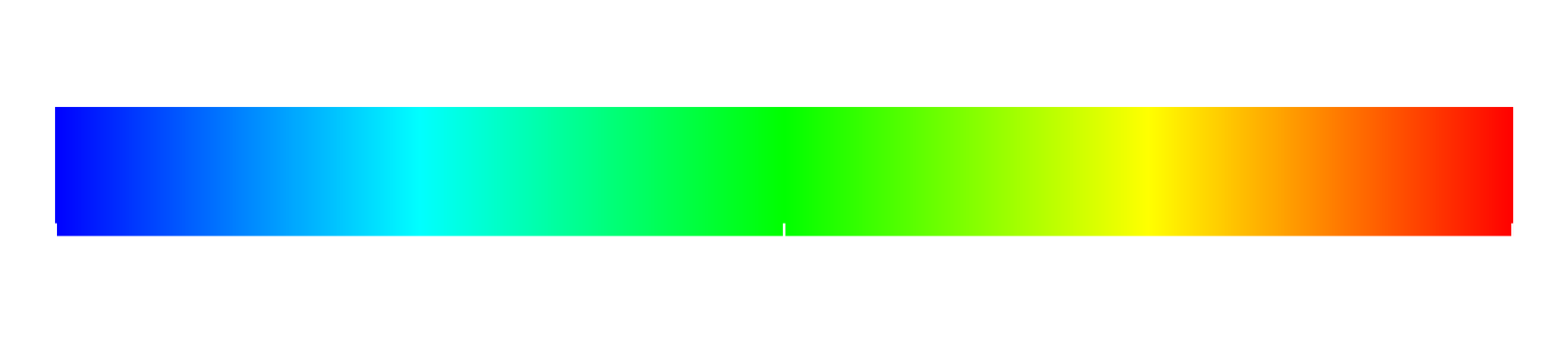}}  
             \end{overpic}   
  \caption{$t=0.2\, ms$}
  \end{subfigure}
  \begin{subfigure}[b]{0.32\textwidth}
    \includegraphics[width=\textwidth]{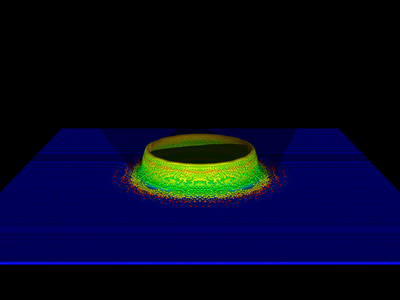}
    \caption{$t=1.7\,ms$}
\end{subfigure}
  \begin{subfigure}[b]{0.32\textwidth}
    \includegraphics[width=\textwidth]{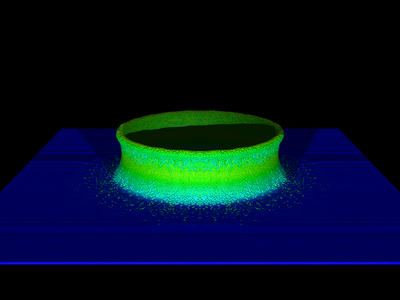}
  \caption{$t=3.45\, ms$}
  \end{subfigure}
  \begin{subfigure}[b]{0.32\textwidth}
    \includegraphics[width=\textwidth]{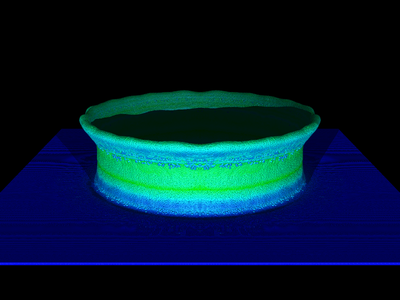}
    \caption{$t=6.2\,ms$}
\end{subfigure}
  \begin{subfigure}[b]{0.32\textwidth}
    \includegraphics[width=\textwidth]{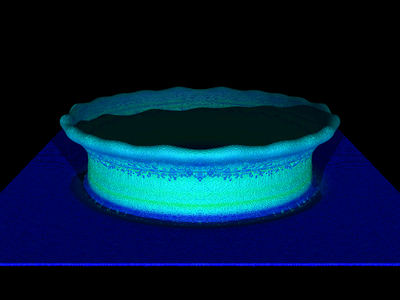}
  \caption{$t=9\,ms$}
  \end{subfigure}
  \begin{subfigure}[b]{0.32\textwidth}
    \includegraphics[width=\textwidth]{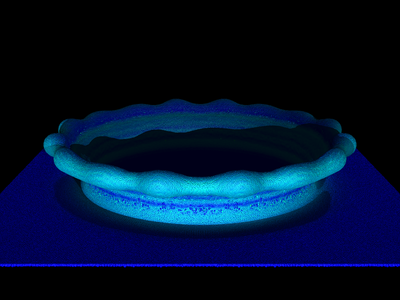}
    \caption{$t=14.9\,ms$}
\end{subfigure}
    \caption{Impact of a droplet on a liquid film with a drop-Weber number, $We_{\textrm{drop}}=273$ and a film-Weber number, $We_{\textrm{film}}=36.5$. The fluid particles are colored by magnitude of velocity, legend on the first image, a. }
    \label{splashimages}
\end{figure}

\begin{figure}[htb]
  \begin{center}
    \includegraphics[width=5in]{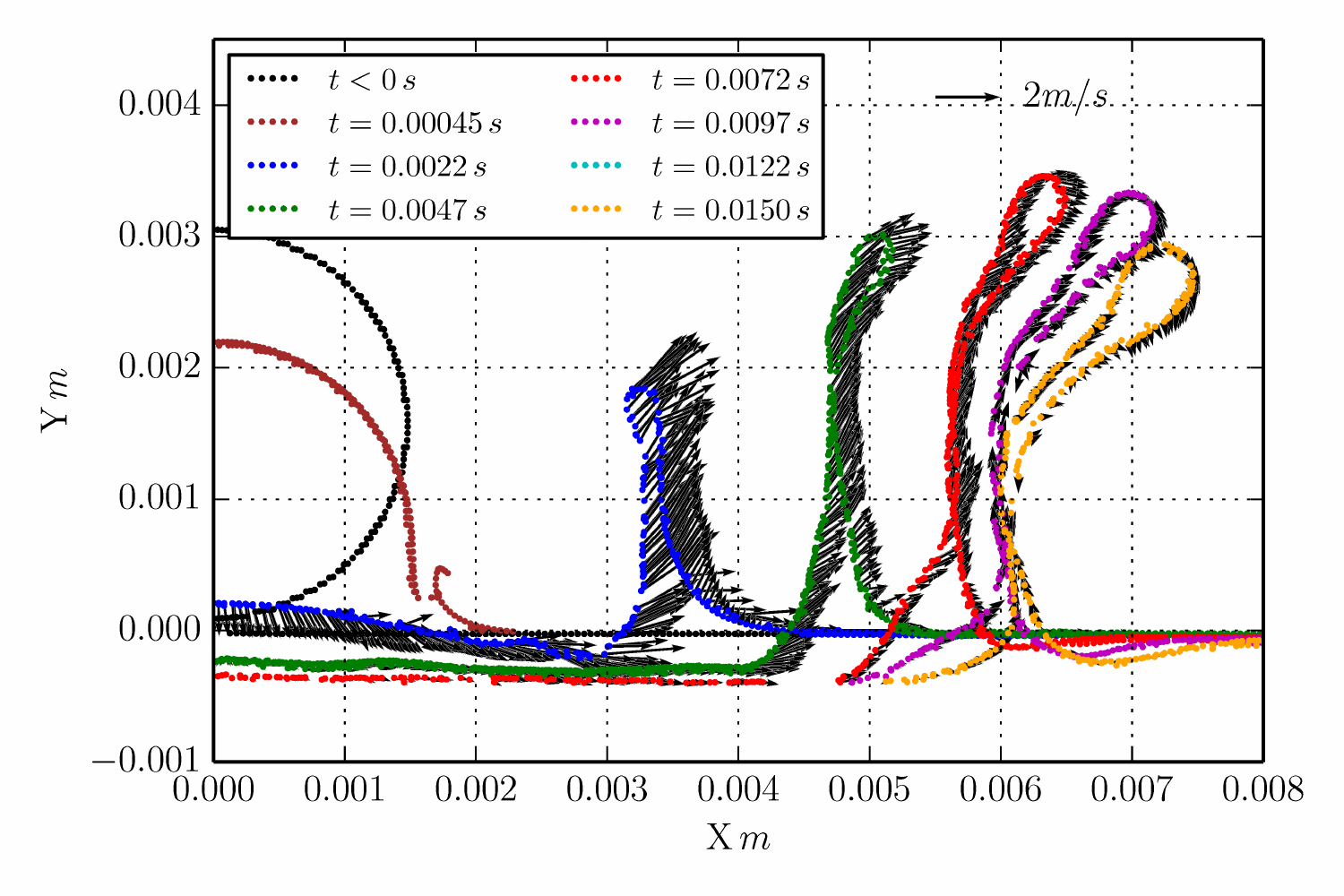}
    \caption{Cross section profile of the splash at different time instances. The grey arrows 
    show the velocity direction of each particle on the surface.}
    \label{splashprof}
  \end{center}
\end{figure}

\begin{figure}[htb]
  \begin{center}
    \begin{subfigure}[b]{0.49\textwidth}
      \includegraphics[width=\textwidth]{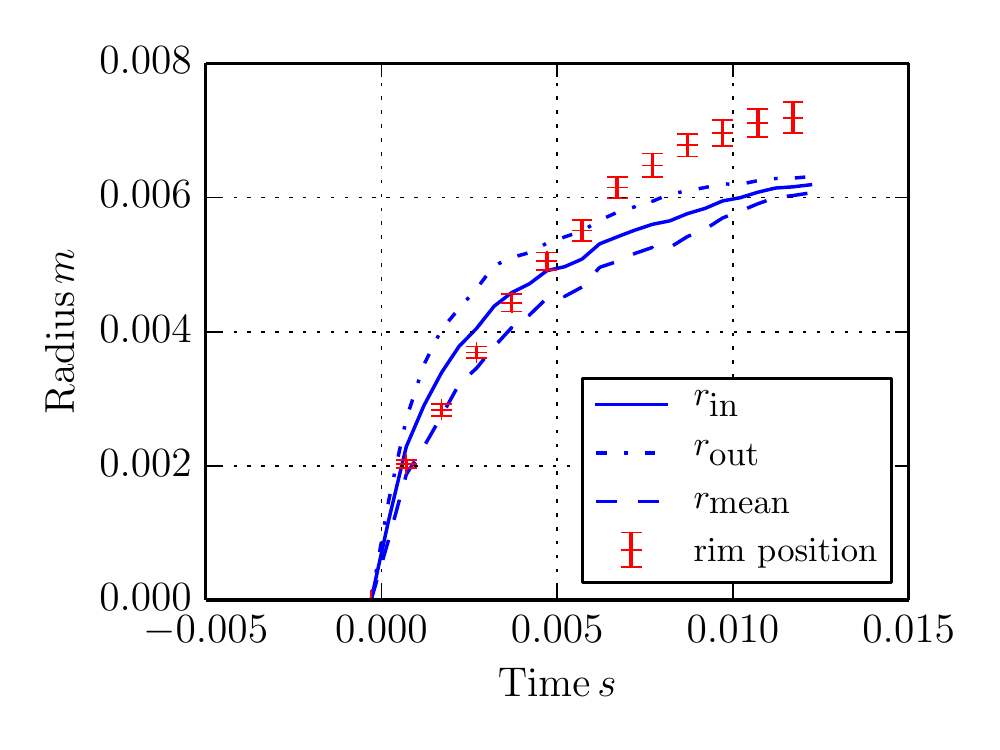}
      \caption{Evolution of crown rim and base}
      \label{fig:height_time}
    \end{subfigure}
    \begin{subfigure}[b]{0.49\textwidth}
      \includegraphics[width=\textwidth]{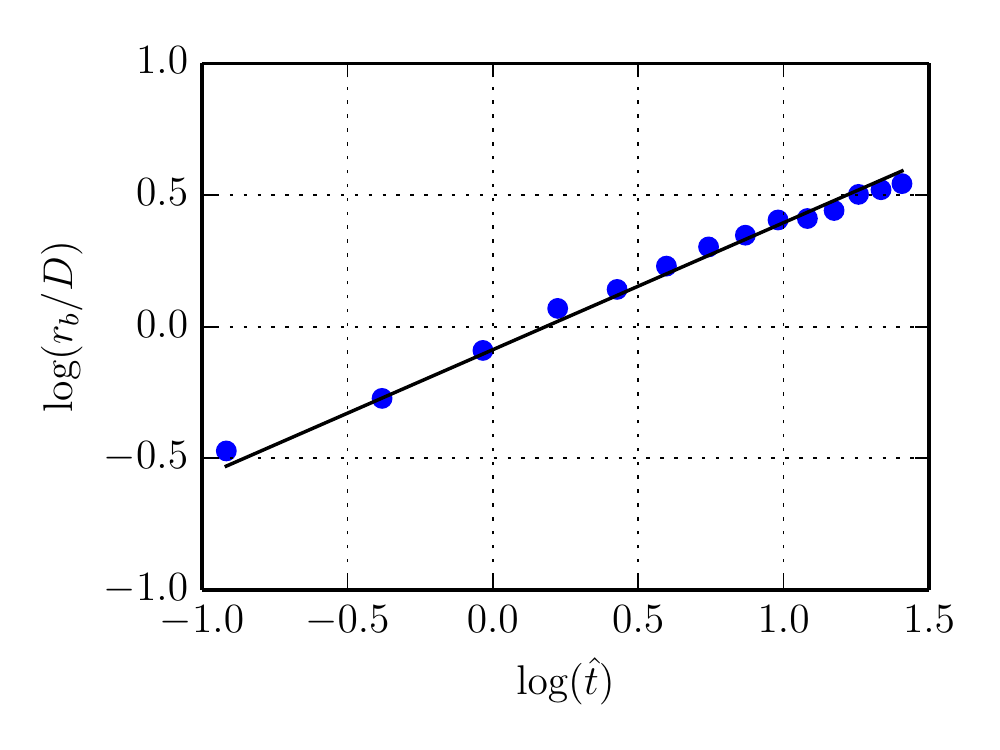}
    \caption{Evolution of base of crown}
      \label{fig:rim_time}
    \end{subfigure}
    \caption{Time evolution of the splash crown, measured at one vertical plane: $r_\text{in}$--inner radius of base of crown, $r_\text{out}$--outer radius of base, $r_\text{mean}$--mean radius of base of the crown.(a) The center and position of the rim at edge of the crown is shown by red markers to scale. (b) Power law dependence of non-dimensional base radius on non-dimensional time. }
  \end{center}
\end{figure}
\begin{figure}[htb!]
\begin{center}
\includegraphics[width=0.8\textwidth]{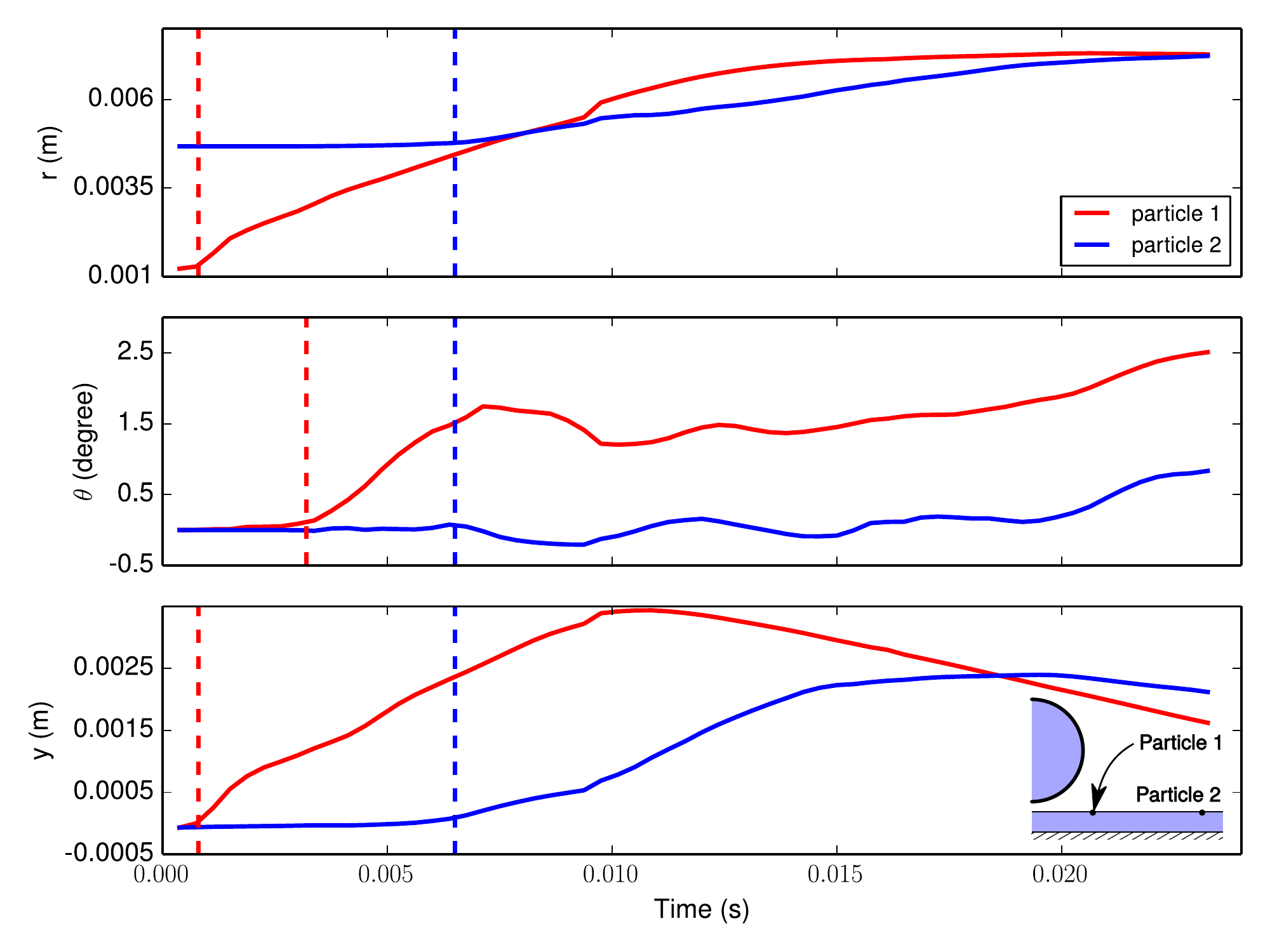}
\caption{Trajectory of two material points in the liquid film }
\label{twoparticles}
\end{center}
\end{figure}

We simulate the splash of a droplet on a film of the same liquid using the introduced method, after
validations. The famous photo of the milk crown by Edgerton and Killian 
\cite{edgerton1954flash} has inspired a number of research works to understand the instability that leads to formation of a crown after a drop of milk impacts a thin liquid film.
 A thin liquid sheet, ejects immediately from the 
neck of the impact. Shortly after a uniform splash sheet is formed, surface tension 
effects begin to dominate and the phenomenon becomes non axisymmetric due to instabilities. The initial part of the splash is driven by mostly inertial effects, 
and this was studied by axisymmetric simulations \cite{josserand} and also by free surface SPH simulations \cite{nair2014improved}. 

The non dimensional parameters---Weber Number ($ We_{drop} =\rho v^{*2} d/\sigma$) and inertia ratio ($\alpha = d/h^*$, where 
$d$ is the diameter of the drop and $h^*$ is the thickness of the liquid film)---influence the formation of secondary droplets\cite{krechetnikov2009crown}. The splashing regime encompasses a variety 
of different morphologies depending on the size and distribution of these droplets \cite{krechetnikov2009crown}. Whether this regime can be explained by a instability mechanism or multiple mechanisms remains an open question \cite{agbaglah2014growth,fullana1999stability} today.

We simulate a $3\,mm$ diameter drop splashing on a liquid film impacting at a velocity of $1.8$ m/s on a liquid sheet of height $0.4\, mm$, set as a square trough periodic in both the horizontal directions. The surface tension coefficient for the fluid is set to $0.03505$ N$/$m, corresponding to 
experiments with milk drops in \cite{krechetnikov2009crown}. This corresponds to a Weber number of $273$. Figure \ref{splashimages} shows the simulation at different time instances, colored by velocity magnitude. In fig. \ref{splashimages}d, the onset of an azimuthal undulation is clearly seen. This undulation clearly compares to the splash crown fingers 
in experiments: 20 crests and troughs are seen in fig. \ref{splashimages} in agreement with the number of sub droplets observed
in experiments of milk crown splash (see fig. 2 in \cite{krechetnikov2009crown}) for the corresponding Weber and inertia numbers.

Lattice Boltzmann \cite{ming2014lattice} and DNS 
\cite{rieber1999numerical}  simulations have been performed 
with the  imposition of  a perturbation
of a given wave number or a Gaussian noise, respectively on the liquid film, to trigger 
the instability.  In our simulations 
no perturbations are imposed, but the noise inherent in SPH method (see fig. \ref{temperature}) 
triggers the crown forming instability. Splash crown simulations in \cite{aly2013modelling} also make use of ISPH. The undulation on the crown, however, is not discernible from their images, which could be due to the fact that free surface particles were identified to impose the Dirichlet boundary condition.
In contrast, the 
free surface in our simulation is smooth and the features of the splash are discernible. 

Figure \ref{splashprof} shows the thickening of the rim of the crown followed by folding of the rim as the expansion of the crown comes to a halt. The arrows
indicate the instantaneous velocity of the surface and shows no source of circulation. The temporal evolution of the crown is shown in fig. \ref{fig:height_time}. The red markers show the position of the rim of the splash wall over-shooting its base as the base of the splash stops expanding.  Figure \ref{fig:rim_time} show the non-dimensionalized radius of base of the splash wall having a power law dependence on the non-dimensionalized time. The theoretical results by \cite{yarin1995impact} predict the following power law relation:
\begin{equation}
\frac{r_b}{D} = k \hat{t}^{1/2}. 
\end{equation}
Here, $r_b$ is the radius of the splash wall at its base, $D$ is the diameter of the drop, $k$ is a constant that depends on the velocity distribution inside the wall of the splash and $\hat{t}=tV/D$ is the non-dimensional time, where $V$ is the impact velocity of the drop. The slope of line in fig. \ref{fig:rim_time} has a value 0.482 which closely corresponds to the square root dependence on time.

The onset of the instability in the splash can be observed as we follow two particles
in different location on the initial liquid film as shown in fig. \ref{twoparticles}.
Particle 1 and 2 are chosen as shown in the inset sketch, and are followed as the 
splash front propagates through them. In the top part of the figure showing
radial displacement, both particles are radially displaced by the splash front,
as soon as the wave touches them. In the middle of fig. \ref{twoparticles},  particle 1 is at the 
azimuthal location ($\theta = 0$) for a brief period of time until the instance 
marked by the red dashed line, after which the particle is azimuthally displaced ($\theta \neq 0$). Particle 2, however is azimuthally displaced right when the splash front reaches it. The 
vertical displacement also follows trend of the radial displacement. There is therefore
a `throw' of the free surface radially and vertically as a wave propagates through it. At some location between the initial positions of particle 1 and 2 the axisymmetry of the wave breaks due to instabilities in the splash crown. This observation is presented to show the advantage a Lagrangian method in tracking instabilities.

\section{Dynamic liquid bridge breakage}
\begin{figure}[htb!]
  \begin{subfigure}[b]{\textwidth}
  \begin{subfigure}[b]{0.24\textwidth}
    \includegraphics[width=\textwidth]{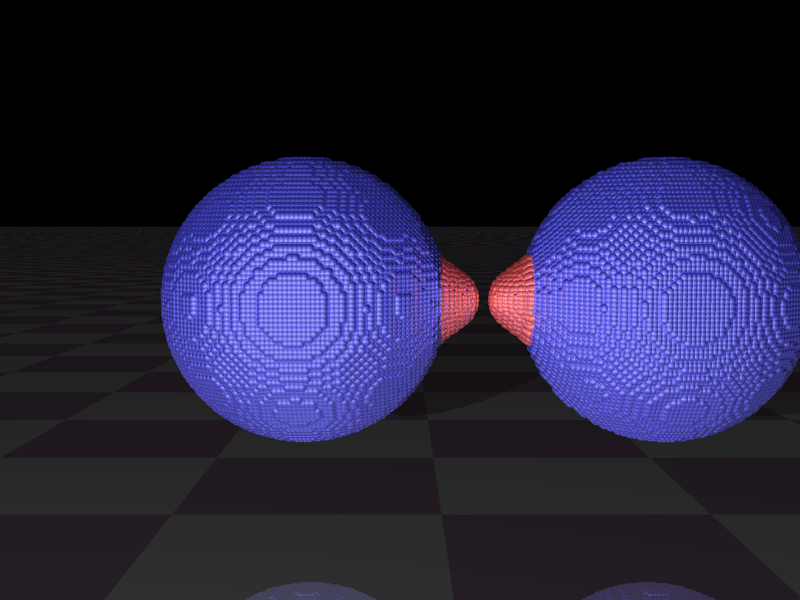}
  \end{subfigure}
  \begin{subfigure}[b]{0.24\textwidth}
    \includegraphics[width=\textwidth]{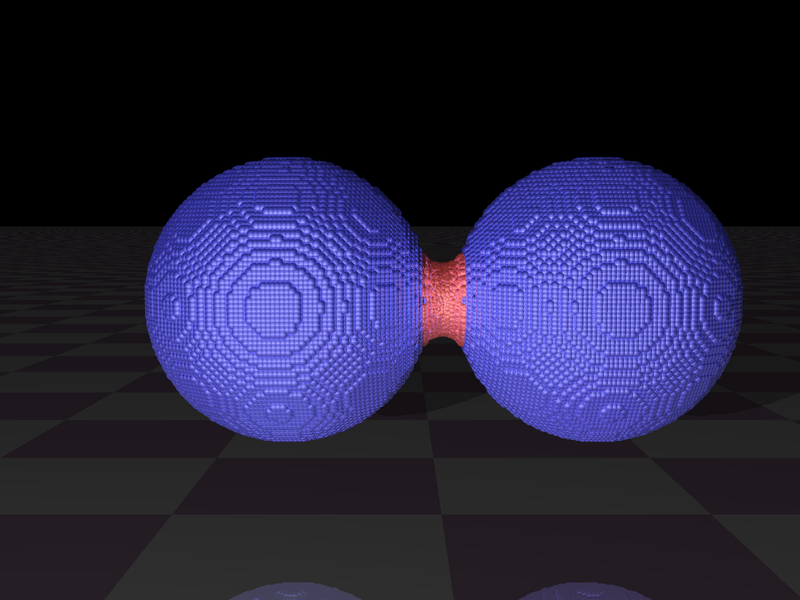}
  \end{subfigure}
  \begin{subfigure}[b]{0.24\textwidth}
    \includegraphics[width=\textwidth]{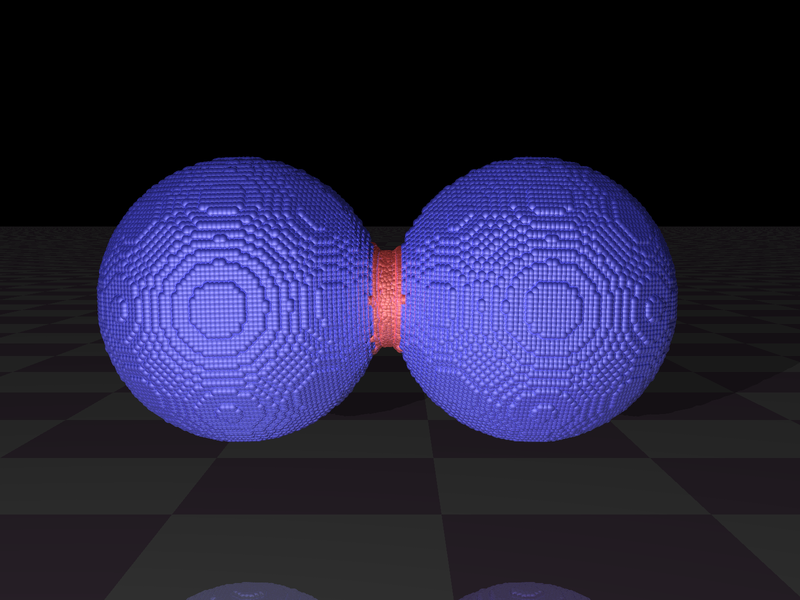}
  \end{subfigure}
  \begin{subfigure}[b]{0.24\textwidth}
    \includegraphics[width=\textwidth]{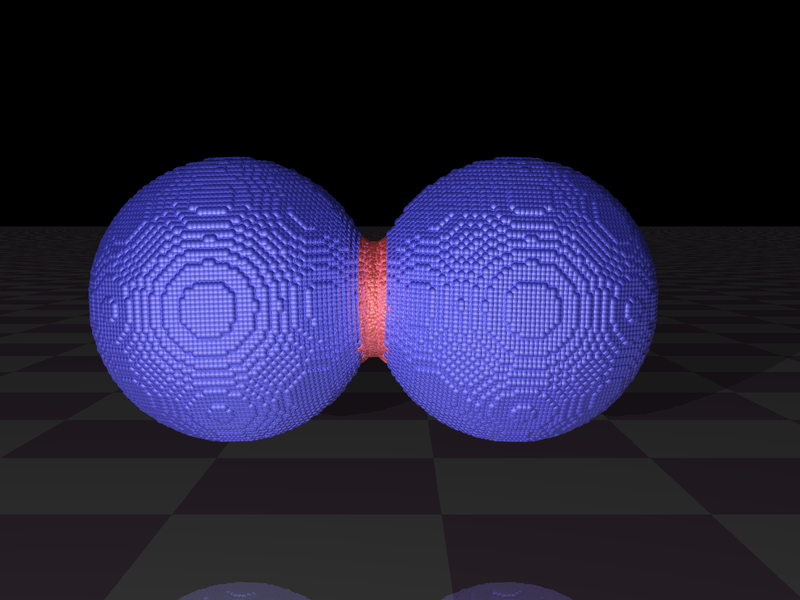}
  \end{subfigure}
    \caption{$0.5\, m/s$}
    \label{agglo}
  \end{subfigure}
  \begin{subfigure}[b]{\textwidth}
  \begin{subfigure}[b]{0.24\textwidth}
    \includegraphics[width=\textwidth]{{visit_fast_1}.png}
  \end{subfigure}
  \begin{subfigure}[b]{0.24\textwidth}
    \includegraphics[width=\textwidth]{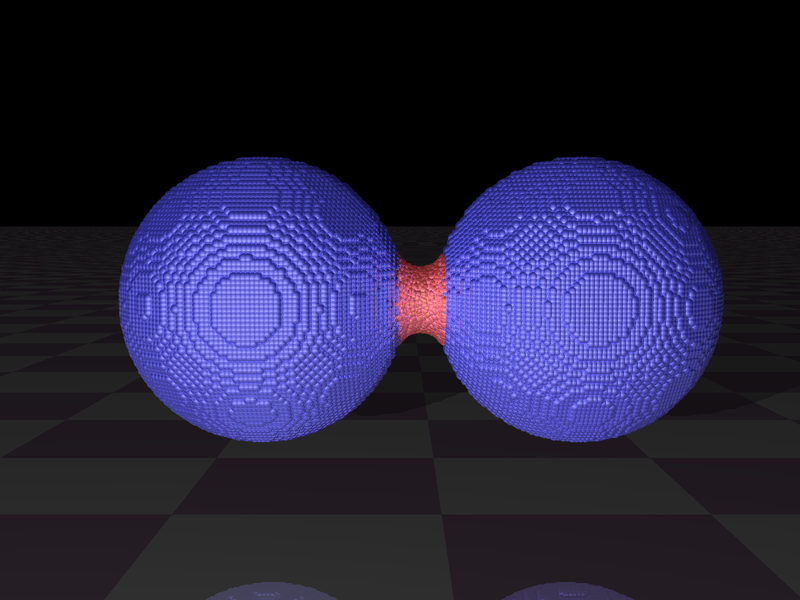}
  \end{subfigure}
  \begin{subfigure}[b]{0.24\textwidth}
    \includegraphics[width=\textwidth]{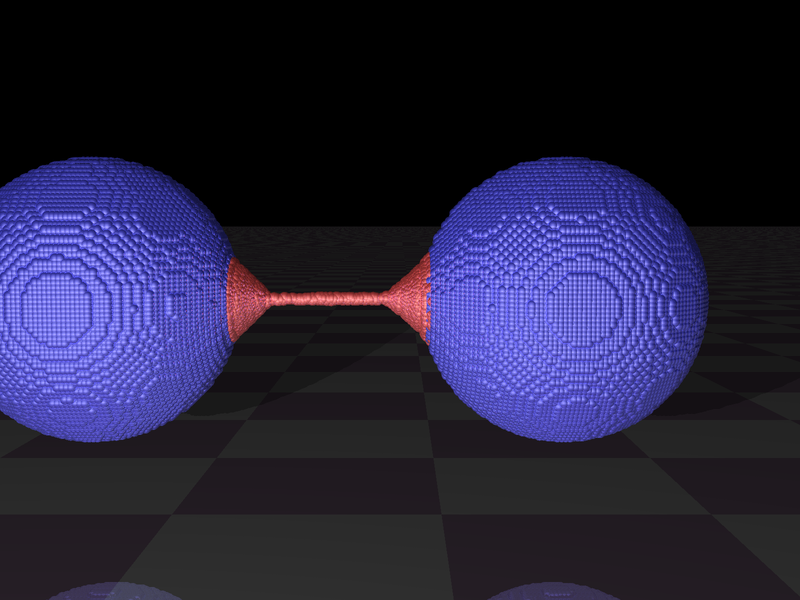}
  \end{subfigure}
  \begin{subfigure}[b]{0.24\textwidth}
    \includegraphics[width=\textwidth]{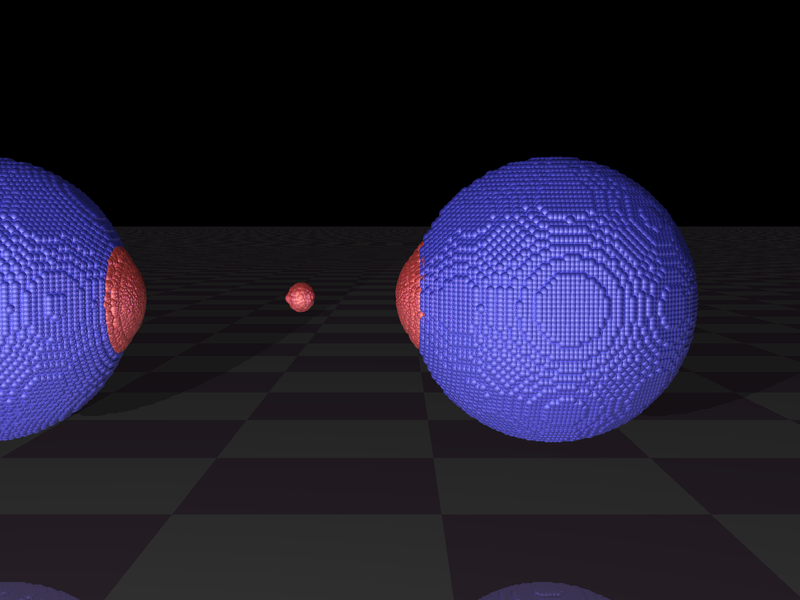}
  \end{subfigure}
    \caption{$5\, m/s$}
    \label{rupture}
  \end{subfigure}
  \caption{Time instances of collision of two wet solid spheres, for different
  approach velocities.}
  \label{dynamiclb}
\end{figure}

\begin{figure}[htb!]
  \begin{subfigure}[b]{0.48\textwidth}
    \includegraphics[width=\textwidth]{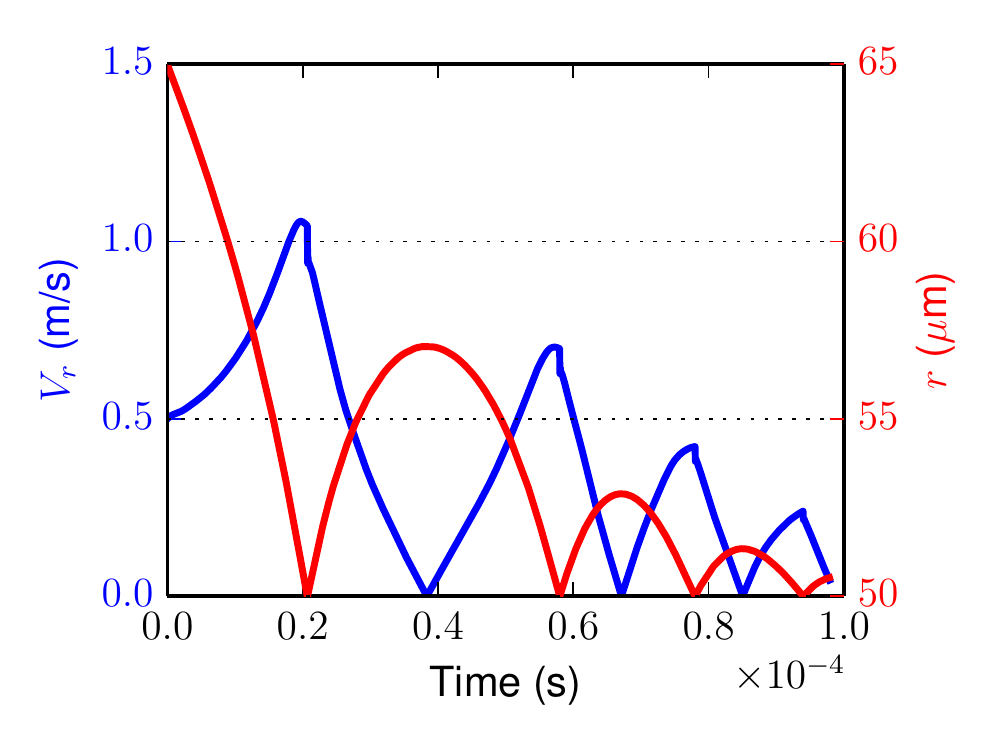}
    \caption{$0.5$ m$/$s}
    \label{coll_0_5}
  \end{subfigure}
  \begin{subfigure}[b]{0.48\textwidth}
    \includegraphics[width=\textwidth]{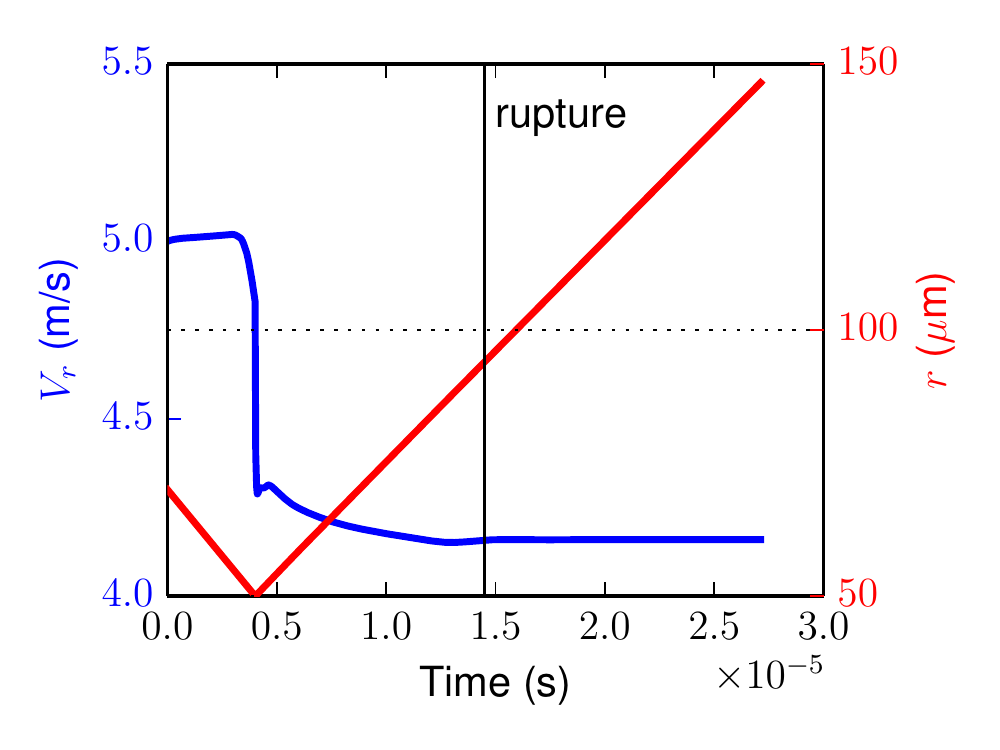}
    \caption{$5$ m$/$s}
    \label{coll_5}
  \end{subfigure}
  \caption{Relative velocity (left y-axis) and distance (right y-axis) between colliding wet solid spheres for different approach speeds.}
  \label{distancevel}
\end{figure}
In wet granulation processes, agglomeration of granules is a critical step 
for successful simulations using macroscopic simulation methods such as Discrete
Element Method (DEM) \cite{tsunazawa2016contact}
or Monte Carlo methods \cite{hussain2013modeling}. The analytical criteria used to decide whether capillary
bridges will be formed during collision of wet particles for given liquid volume
and approach speeds is given by quasi-static considerations. 
Depending on the dominance of viscous or surface tension effects in the 
binding liquid, a specific model is used. For example, for viscous dominant 
binders the following critical velocity for agglomeration is given \cite{ennis1991microlevel}:
\begin{equation}
  v_c = \frac{3\pi\mu R_p^2}{2M_p}\left( 1 +\frac{1}{e} \right)\ln \left( \frac{y}{y_a}\right),
  \label{stokes}
\end{equation}
where $y$ and $y_a$ are the thickness of the liquid layer on the grain and 
the surface asperity on the grain, respectively. $R_p$, $M_p$ and $e$ are the 
radius, mass and restitution coefficient of the grains, respectively. 
Other class of analytical models assume a surface tension dominance to compute
the force between the grains for different static configurations of the liquid 
bridge, for example a pendular shape \cite{pitois2001rupture}. In reality,
the bridge deforms quite rapidly and inertial forces compete with surface tension forces
deforming the interface non-linearly and a general analytical solution to the problem is difficult.

We performed a range of simulations of colliding wet spheres to find the 
critical velocity at which the particles stop agglomerating. This is compared
to results from the purely viscous analytical model given in eq.\ref{stokes} 
and a numerical result using a grid based method \cite{kan2015numerical}. 
Two solid spheres of diameter 50 $\mu$m are considered, with a  fluid drop of 
total volume $1072\, \mu$m$^3$ divided into two drops and placed as sessile drops on both the spheres. The viscosity and surface
tension of the liquid are 0.001 Nsm$^{-2}$ and 0.071 N/m respectively. These
simulation parameters are similar to those used in \cite{kan2015numerical} for a 
similar study of dynamic liquid bridges. One of the 
particles is launched at the other; upon contact of the liquid surfaces the drops coalesce and a liquid bridge is 
formed. The collision between the rigid solid spheres is assumed to be purely elastic (restitution coefficient, $e=1$) 
and  takes place at a point immersed within the liquid bridge. 

In figure \ref{dynamiclb}, liquid bridge configuration at different time instances
for two different approach velocities are shown. The rupture of liquid bridge following collision of the solid 
spheres is shown in fig. \ref{rupture} for an approach velocity of $5$ m$/$s, which is beyond the critical value
for agglomeration. The liquid bridge thins to form a near cylindrical filament 
before rupture. The filament then breaks and relaxes to a satellite drop. In fig. \ref{agglo}, on the other hand, the liquid bridge sustains itself and causes agglomeration
of the solids. In figure \ref{distancevel}
the distance between the two solid spheres and relative velocity are shown as a function of times. For the smaller 
velocity considered (fig. \ref{coll_0_5}), the bridge sustains and results in the particles bouncing off 
of each other repeatedly while remaining agglomerated. For velocity larger than critical value (fig. \ref{coll_5}), the bridge
ruptures and particles depart at constant velocities following rupture. A series
of simulations were performed and a value of 1.1 m$/$s is found to be the critical 
velocity for agglomeration of the solid spheres. This value is about an order of 
magnitude higher than that predicted by the analytical model given by eq. \ref{stokes}:
a value of 0.16 m$/$s. In a similar study using a constrained interpolation method with a VOF model for interface predicted a critical velocity of 2.6 m/s \cite{kan2015numerical}. The reason for this discrepancy should depend on the resolution of the interface near the rupture and requires verification with experiments yet to be published. However a clear underprediction of critical velocity by a widely used analytical model is evident.  With a series of numerical experiments very accurate 
models for the micromechanics of collision/agglomeration can therefore be 
developed for different regimes of viscous and surface tension forces, using 
the introduced ISPH method.

\section{Conclusion}
Incompressible Smoothed Particle Hydrodynamics (ISPH), 
a more accurate and robust variant of the SPH method is 
improved in scope to include freesurface surface tension 
and wetting problems. The competence of this improved meshless
method in studying complex flows encountered at small scales
is demonstrated through three dimensional simulations of 
dynamic capillary phenomena. The unexplained artificial viscosity
due to inter particle forces needs further analysis. 

\newpage

\clearpage
\newpage
\bibliographystyle{elsart-harv}


\end{document}